\documentclass[useAMS,usenatbib]{mn2e}
\usepackage{epsfig}
\usepackage{graphicx}
\usepackage{times}
\usepackage{color}
\usepackage{amssymb}
\usepackage{array}
\usepackage{booktabs}
\usepackage[T1]{fontenc}
\usepackage{amsmath}

\usepackage{caption}

\def\cm3{cm$^{-3}$}
\def\kms{km~s$^{-1}$}
\def\msunyr{M$_{\odot}$\,yr$^{-1}$}

\def\msun{M$_{\odot}$}

\def\beq{\begin{equation}}
\def\eeq{\end{equation}}

\def\lesssim{\mathrel{\hbox{\rlap{\hbox{\lower4pt\hbox{$\sim$}}}\hbox{$<$}}}}
\def\gtrsim{\mathrel{\hbox{\rlap{\hbox{\lower4pt\hbox{$\sim$}}}\hbox{$>$}}}}

\def\lesssim{\mathrel{\hbox{\rlap{\hbox{\lower4pt\hbox{$\sim$}}}\hbox{$<$}}}}
\def\gtrsim{\mathrel{\hbox{\rlap{\hbox{\lower4pt\hbox{$\sim$}}}\hbox{$>$}}}}

\newcolumntype{L}{>{\centering\arraybackslash}m{1.9cm}}

\def\heracles{{\sc heracles}}

\newcommand{\iso}[2]{\ensuremath{^{#1}\rm{#2}}}

\def\aj{AJ}
\def\pasp{PASP}
\def\apj{ApJ}

\def\aap{A\&A}
\def\araa{ARA\&A}

\def\mnras{MNRAS}
\def\nat{Nature}

\def\jqsrt{JQSRT}

\title[Super-luminous type IIn SNe]{Two-dimensional radiation-hydrodynamics simulations
of super-luminous interacting supernovae of type IIn}
\author[Alkiviadis Vlasis et al.]
{Alkiviadis Vlasis,$^{1}$ Luc Dessart,$^{1}$ and Edouard Audit$^{2}$\\ \\
$^{1}$: Laboratoire Lagrange, UMR7293, Universit\'e Nice Sophia-Antipolis, CNRS,
Observatoire de la C\^{o}te d'Azur, 06300 Nice, France. \\
$^{2}$: Maison de la Simulation, CEA, CNRS, Universit\'e Paris-Sud, UVSQ, Universit\'e Paris-Saclay, 91191, Gif-sur-Yvette, France. \\
}

\begin{document}

\date{Accepted . Received }

\pagerange{\pageref{firstpage}--\pageref{lastpage}} \pubyear{2016}

\maketitle

\label{firstpage}

\begin{abstract}
Some interacting supernovae (SNe) of type IIn show a sizeable continuum polarisation suggestive of
a large scale asymmetry in  the circumstellar  medium (CSM) and/or the SN ejecta.
Here, we extend the recent work of Dessart et al. on super-luminous SNe IIn and perform
axially-symmetric (i.e., 2D) multi-group radiation hydrodynamics
simulations to explore the impact of an imposed large scale density asymmetry.
When the CSM is asymmetric, the latitudinal variation of the radial optical depth $\tau$ introduces
a strong flux redistribution from the higher-density CSM regions, where the shock luminosity is larger,
towards the lower-density CSM regions where photons escape more freely --- this redistribution
ceases when $\tau\lesssim$\,1.
Along directions where the CSM density is larger, the shock deceleration is stronger and its progression
slower, producing a non-spherical cold-dense shell (CDS). For an oblate CSM density distribution,
the photosphere (CDS) has an oblate (prolate) morphology when $\tau\gtrsim$\,1.
When the CSM is symmetric and the ejecta asymmetric, the flux redistribution within the CSM now tends to
damp the latitudinal variation of the luminosity at the shock. It then requires a larger ejecta asymmetry
to produce a sizeable latitudinal variation in the emergent flux.
When the interaction is between a SN ejecta and a relic disk, the luminosity boost at early times
scales with the disk opening angle -- forming a super-luminous SN IIn
this way requires an unrealistically thick disk. In contrast, interaction with a  disk of modest
thickness/mass can yield a power that rivals radioactive decay of a standard SN II at nebular times.
\end{abstract}


\section{Introduction}
\label{sect_intro}

SNe IIn represent $\approx$\,10\% of all core-collapse SNe \citep{Li2011}, and are characterised by
the presence of narrow spectral lines at discovery \citep{Schlegel1990}.
Some SNe IIn exhibit a luminosity that far exceeds that of standard, i.e., non-interacting, SNe.
This luminosity boost is believed to stem from the extraction of ejecta kinetic energy during an interaction
with a dense, extended, and massive circumstellar medium (CSM) formed by the progenitor star during its evolution.
The origin of such a massive CSM is debated.
It may be produced through violent envelope pulsations, e.g., in connection to the pair-production instability in very massive
stars \citep{Barkat1967,Woosley2007}.
Another means is through a super-Eddington wind mass loss \citep{Davidson_Humphreys_1997, Owocki2004}.
A nuclear flash a few years before core collapse also seems a robust mechanism, with a natural synchronisation
for producing an interaction, but it probably operates over a narrow mass range
(around 9-11\,\msun; \citealt{WH15}; \citealt{chugai_15}; \citealt{D16})

Numerical simulations have been used to study the properties of interacting SNe.
\citet{van_marle_etal_10} studied the hydrodynamics of SNe IIn, treating the radiation
through a parametrised cooling term in the energy equation. This appoximation is suitable
for optically-thin conditions.
However, in the context of super-luminous SNe IIn, the CSM is massive, dense, and extended (otherwise
little kinetic energy can be extracted from the SN ejecta to boost the luminosity) and hence, once ionised, the CSM
is optically thick. Consequently, the radiation emerging from super-luminous SNe IIn is subject to strong optical
effects which require a modelling with multi-group radiation hydrodynamics
(\citealt{Chugai2004, Woosley2007, moriya_etal_13b, Whalen2013,Dessart_2015}, hereafter D15).

In D15, we presented such simulations. The reference model X of D15, whose initial parameters are also used
in the present work, matches favourably the light curve and spectral evolution of the super-luminous type IIn SN\,2010jl.
This model consists of a 10\,\msun\ 10$^{51}$\,erg ejecta ramming
into a $\sim$\,3\,\msun\ CSM moving at 100\,\kms\ and extending from 10$^{15}$ to about 10$^{16}$\,cm.
The basic features of this interaction model are the following (see D15 for details).
A strong shock forms at the ejecta/CSM interface and releases a large luminosity, which, under the influence of
absorption, scattering, and emission,  crosses the CSM on a time scale of a week.
This ``radiative precursor'' ionises this extended CSM, which in the process becomes optically thick.
The subsequent shock luminosity then has to diffuse through the CSM, producing a bell-shape light curve morphology.
Optical-depth effects last for as long as the CSM above the shock remains optically thick, which depends
on the shock propagation speed, the CSM mass and extent etc --- in the case of SN\,2010jl, the optically-thick
phase lasts nearly a year.
As the shock progresses outwards, a radially-confined cold dense shell (CDS) forms (bounded by the forward and reverse shocks)
and grows in mass,  eventually sweeping through the entire CSM.
The high brightness phase of super-luminous SNe like 2010jl persists as long as there is a dense CSM to interact with.
In D15, we find that the CSM is not massive and extended enough to decelerate completely the inner shell material so
that the conversion efficiency of ejecta kinetic energy to emergent radiation is $\sim$30\% once the interaction has died out.
This corresponds to a total time-integrated luminosity of about $3 \times 10^{50}$\,erg, which is 30 times larger than for
a typical SN II.
The observed spectral evolution of SN\,2010jl corroborates the inferences from the light curve evolution.
At early times, when the CSM is optically thick, spectral lines are symmetric with narrow cores and extended wings
produced by electron scattering within the optically-thick CSM.
As time progresses and the CSM optical depth above the CDS abates, line profiles start to show a Doppler-broadened
component and a blue-shift of peak emission, reflecting the growing contribution from the fast moving CDS.
At late times, most of the radiation arises from the CDS and the line profiles are symmetric and Doppler broadened.

Interestingly, SN\,2010jl shows a high level of (intrinsic) continuum polarisation \citep{Patat_2011}, which suggests
that the distribution of the flux and/or the material is not uniform on the plane of the sky
\citep{brown_77,ss82,hoeflich_91,Dessart_2011}.
D15 post-processed their 1-D radiative transfer simulation by imposing a large scale asphericity and
found that an oblate/prolate morphology with a pole-to-equator density ratio of 2-3 could explain
the observed level of polarisation at bolometric maximum.
Other SNe IIn have also shown significant polarisation, suggestive of an asymmetric ejecta/CSM configuration
\citep{leonard_98S_00, Hoffman2008}.
Understanding the properties of the CSM, how it formed, and how it departs from spherical symmetry is of considerable
interest.

Observations reveal that the CSM around massive stars is often aspherically distributed.
The Homunculus nebula around the star $\eta$ Carinae has a bipolar morphology and contains 10-20\,\msun\ of material, probably
ejected over a timescale of only a decade \citep{Davidson_Humphreys_1997,Humphreys1999_etacar}.
The red-supergiant (RSG) star VY CMa exhibits an asymmetric, clumpy, and dense wind \citep{Smith2001,Wittk2012}.
The RSG star Betelgeuse also shows signs of asymmetric, albeit weaker, mass loss
\citep{Smith2009_Betelgeuse,Ohnaka2011_Betelgeuse}.
In some cases, the CSM may be a relic disk, which prevailed through the life of the massive star.
Speculations on the existence of (and interaction with) such disks have been made for SN\,1997eg \citep{Hoffman2008}.
More recently, \citet{Metzger2010} studied the interaction of a SN ejecta with such a massive (1-10\,\msun) relic disk
and argued that super-luminous SNe IIn could be produced through this scenario.
\citet{Smith2015_PTF} invoked the ejecta/disk interaction as a means to  power the late time luminosity and explain
the spectral line profile morphology of  SN PTF11iqb.

An asymmetric CSM does not preclude the possibility that the ejecta produced by
the terminal (or non terminal) explosion of the star is also asymmetric.
The persistent triple-peaked H$\alpha$ observed in SN\,2010jp is suggestive of a bipolar explosion
in a Type II SN \citep{smith_10jp_12}.
The Type II-P SN\,2004dj reveals a large increase in continuum polarisation as the inner ejecta is revealed
at the end of the plateau phase, suggesting the explosion itself was asymmetric \citep{Leonard2006}.
More recent spectropolarimetric observations of SNe II confirm this finding, but also emphasise the diversity
of the measured continuum polarisation (e.g., sizeable polarisation prior to, or sometimes only at the onset of,
the nebular phase) and the processes at their origin
(large scale asymmetry, $^{56}$Ni fingers etc; see \citealt{leonard_15} for discussion).

So far, radiation hydrodynamic simulations of SNe IIn have assumed spherical symmetry.
Here, we remedy this shortcoming by performing axially-symmetric (2-D) multi-group radiation hydrodynamic simulations
of super-luminous SNe IIn, breaking the spherical symmetry of the initial interaction model by introducing a latitudinal scaling
in the density distribution.
Just like in D15, we focus on super-luminous SNe IIn because they correspond to interactions involving a CSM of a large mass,
and therefore of a large optical depth once ionised.
But we now wish to study how the breaking of spherical symmetry impacts both the dynamics of the interaction
and the emergent radiation from the interaction.
We focus on large scale asymmetries using a coarse angular resolution, which is too small
to capture the development of fluid instabilities that might otherwise develop on small scales, such as
the thin-shell instability.
In Section~\ref{sect_radhydro} we give an overview of our numerical setup.
We then proceed in steps by considering interactions in which only the CSM is asymmetric  (Section~\ref{sec_asymcsm}),
in which only the ejecta is asymmetric (Section~\ref{sec_asym_ej}), and finally the case in which a spherically-symmetric
ejecta interacts with a relic disk  (Section~\ref{sec_disk}).
We then present our conclusions in Section~\ref{sec_discussion}.

\begin{table*}
\caption{Summary of initial conditions and key results for ejecta interactions
with an asymmetric CSM. We include the initial pole-to-equator density ratio, as well
as the values at the time of bolometric maximum ($t_{\rm peak}$) for
the pole-to-equator ratio of the radius, density, and flux at the photosphere, at the CDS,
or at $r_{\rm max}$.
\label{tab_csm}
}
\begin{tabular}{l@{\hspace{5mm}}c@{\hspace{4mm}}c@{\hspace{4mm}}c@{\hspace{4mm}}
c@{\hspace{4mm}}c@{\hspace{4mm}}c@{\hspace{4mm}}c@{\hspace{4mm}}
c@{\hspace{4mm}}c@{\hspace{4mm}}c@{\hspace{4mm}}c@{\hspace{4mm}}}
\hline
Model    & CSM &  $A$  & $n$
& $(\rho_{\rm p}/\rho_{\rm e})_{t=0}$
& $t_{\rm peak}$ [d]
& $(r_{\rm p}/r_{\rm e})_{\rm Phot}$
& $(r_{\rm p}/r_{\rm e})_{\rm CDS}$
& $(v_{\rm p}/v_{\rm e})_{\rm CDS}$
& $(\rho_{\rm p}/\rho_{\rm e})_{\rm CDS}$
&$(F_{\rm p}/F_{\rm e})_{r_{\rm max}}$ \\
\hline
CP1           &    prolate         &  2         & 2 &      3      &  26.4 & 1.51 &  0.84   & 0.89   & 2.61       & 0.47    \\
CP2           &    prolate         &  5          & 2 &      6     &   25.4 & 2.27 &  0.78   & 0.82   & 4.64       & 0.30    \\
CO1          &    oblate          &  -0.67   & 2 &      0.33  &  25.5 & 0.63 & 1.08    & 1.13   & 0.35        & 2.07    \\
\hline
\end{tabular}
\end{table*}

\section{Numerical setup}
\label{sect_radhydro}

The simulations presented in this work were performed with the Eulerian radiation hydrodynamics
code \heracles\ \citep{gonzalez_etal_07,vaytet_etal_11}, assuming axially-symmetric (2-D)
configurations and employing the M1 moment method to solve
for the first two angular moments of the specific intensity \citep{m1_model}.
Our approach is also multi-group rather than grey. This matters in simulations of SNe IIn because the radiation and the
gas are strongly out of equilibrium with respect to each other (D15).

Here, we merely extend the simulations of D15 to 2-D. We adopt the same equation of state for the gas (an ideal gas with
an adiabatic exponent of 5/3), the same opacity tables and energy groups, the same uniform composition (the mass fraction of H, He, and Fe are 0.633, 0.36564 and 0.00136).

What differs from D15 is the geometrical domain we study. To keep the computations tractable, we assume equatorial symmetry
and thus simulate one hemisphere in a meridional slice. Using spherical coordinates $(r, \theta)$,
we cover in radius from $r_{\rm min}=$\,5$\times$10$^{13}$\,cm to $r_{\rm max}=$\,1.45$\times$10$^{16}$\,cm,
and in polar angle from 0 to 90\,deg.
Simulations use a uniform radial grid with $n_r=$\,600 points apart from model CP1
which uses 1200 zones. This higher radial resolution reduces a numerical artefact that develops
after about 80 days in model CP1 -- this
artefact completely disappears if we use 2400 zones (see next section).
As in D15, we use an inflow inner boundary (constant velocity gradient through the boundary) and an outflow outer boundary
(constant velocity through the boundary). For the radiation, we impose zero flux at the inner boundary and flow-out at the
outer boundary.
The angular grid is generally uniform and uses $n_\theta=$\,20 angles. Along $\theta=$\,0\,deg (pole) and
$\theta=$\,90\,deg (equator), we adopt reflecting boundary conditions for both the gas and the radiation variables.
For the interaction involving a strongly asymmetric ejecta (model EP2), we increase $n_\theta$ from 20 to 40.
For the ejecta/disk simulations, we adjust the angle grid and adopt a uniform resolution within the disk (20 angles)
and a coarse grid beyond (10 additional angles with a logarithmically increasing spacing up to the polar direction).

We set analytically the velocity, density and temperature of the initial configurations, as described for model X in D15.
In all simulations, the ejecta extends from the inner boundary out to the transition radius $r_{\rm t}=$\,10$^{15}$\,cm
and is in homologous expansion.
Its temperature is the same as that for model X in D15, which is typical of a SN II at $\sim$\,10\,d after explosion.
The total ejecta mass is 9.8\,\msun\ and the total ejecta kinetic energy is 10$^{51}$\,erg.
Beyond $r_{\rm t}$, we fill the grid with different types of CSM (all with a fixed initial temperature of 2000\,K)
including winds (with various angle-dependent mass loss rates, but the same total mass of 2.89\,\msun)
and relic disks (with various total mass and opening angle).

\begin{figure*}
  \epsfig{file=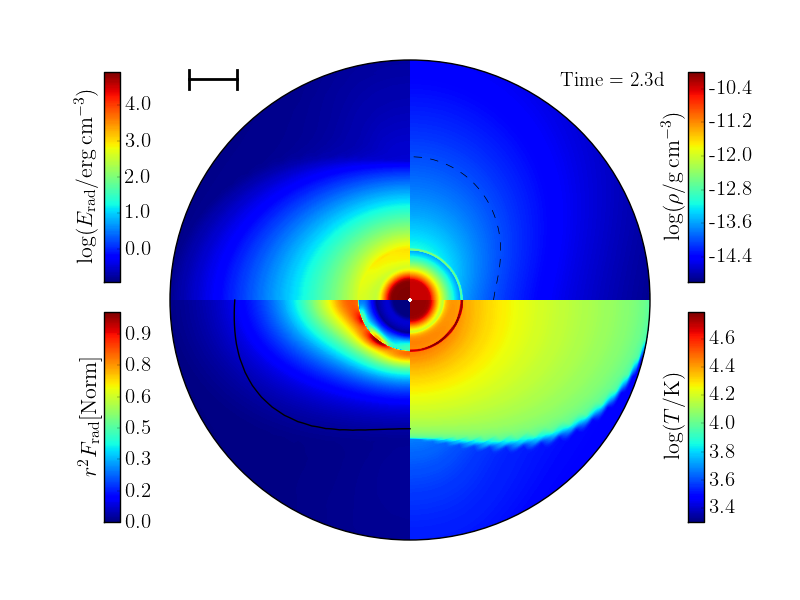,width=8.8cm}
  \epsfig{file=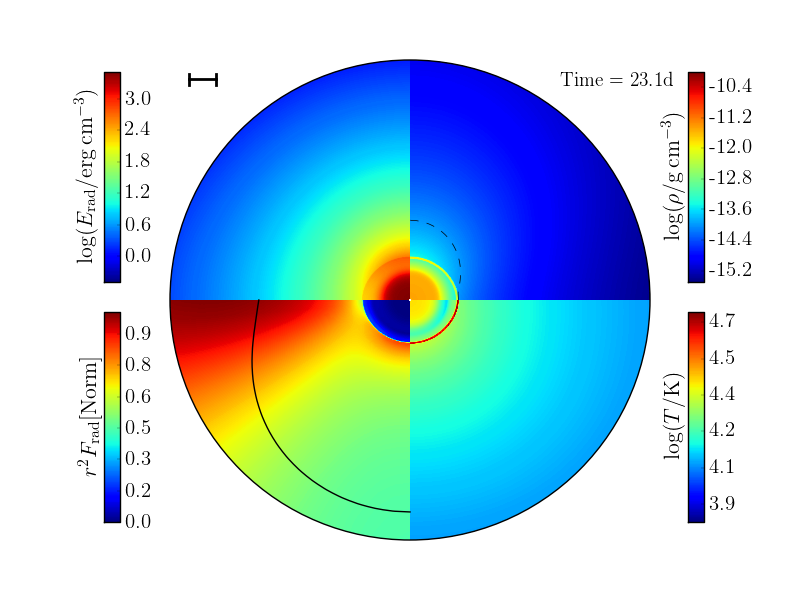,width=8.8cm}
  \epsfig{file=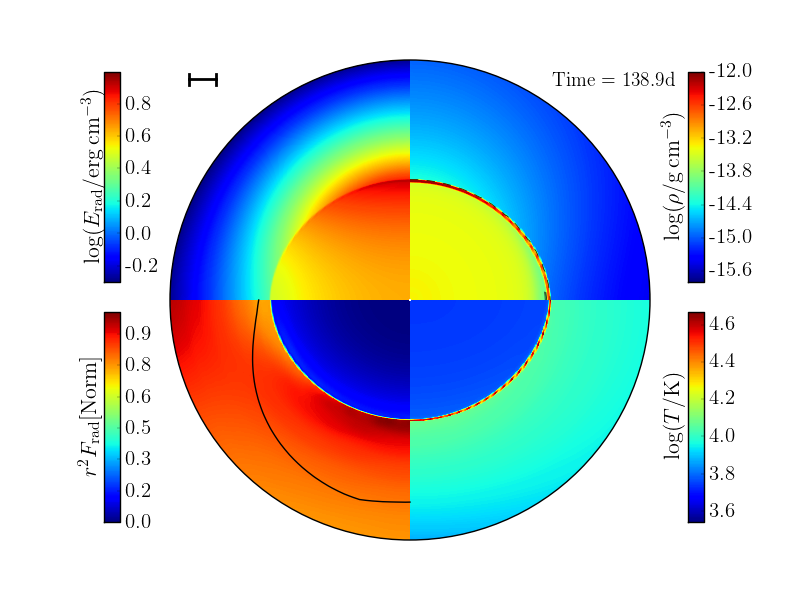,width=8.8cm}
   \epsfig{file=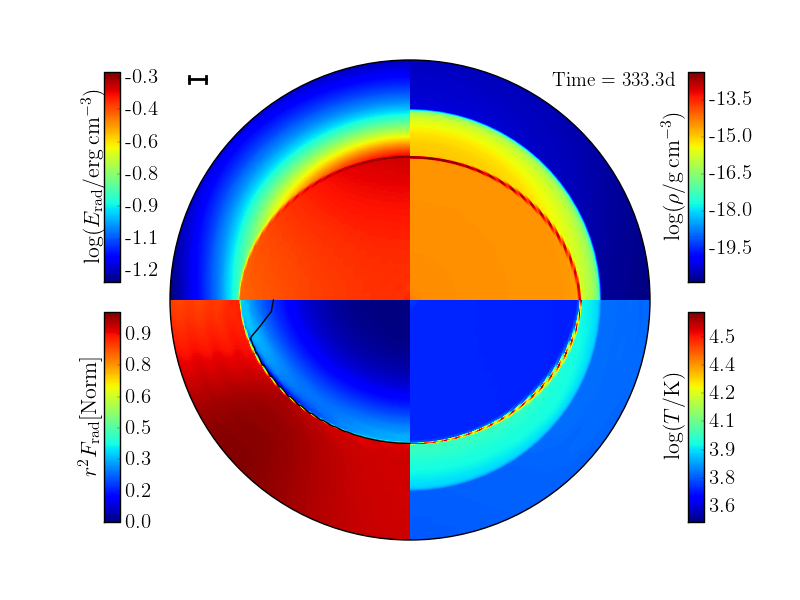,width=8.8cm}
  \caption{
   Top left:
   Dynamical and radiative properties of the reference model CP1 2.3 d after the onset of interaction.
   The four quadrants show the 2-D properties of the density
   (the dashed line in the top row panels corresponds to the 10$^{-14}$\,g\,cm$^{-3}$ contour),
   the temperature, the radiative flux (multiplied by $r^{2}$ and normalised; we overlay the electron-scattering photosphere)
   and the radiative energy.
   Here and in other panels, the length of the thick bar corresponds to a physical scale of 10$^{15}$\,cm.
   At this time, the radiation injected at the shock progressively fills the CSM, raising its temperature, ionisation, and opacity.
   This radiative ``precursor" propagates faster along the lower density equatorial regions.
   Top right:
   Same as top left, but now at 23.1\,d (which corresponds to the epoch of flux maximum along the equator).
   The CSM is now optically thick, the photosphere is more elongated along the poles, and photons emerge
   preferentially through the lower density equatorial regions. The emergent flux has an oblate distribution
   while the density distribution is prolate.
   Bottom left:
   Same as top left, but now at 138.9\,d
   The CDS is now close to the photosphere along the equator.
   Bottom right:
   Same as top left, but now at 333.3\,d. The CDS has overtaken the dense part of the CSM and the
   overlying CSM is optically thin. The CDS, which is less dense along the equator, starts to turn optically thin
   at low latitudes. Compared to early times when the CSM was optically thick, the flux is now greater at higher
   latitudes.
   \label{n2_multi}
      }
\end{figure*}

For the ejecta/wind interaction simulations, we introduce the large-scale asymmetry through a latitudinal scaling of the density.
This is a rather crude modelling of the explosion/eruption/wind at the origin of the asymmetric
CSM and ejecta but this option is chosen for simplicity. Furthermore, much uncertainty surrounds
the production of a massive CSM of the sort presented here, and the explosion of massive stars following
core collapse is not a settled matter.
In practice, given the radial density distribution of model X used in D15, the density $\rho(r,\mu)$ at $r$ and $\mu=\cos \theta$
is given by $\rho(r,\mu) = \rho(r)(1+A\mu^{n})$, where $A$ and $n$ are parameters that allow the user to tune
the pole-to-equator density ratio and the latitudinal gradient of the density.
To ensure that the outer boundary is optically thin, we reduce the wind mass loss rate by a factor of 10$^5$ in the outer CSM (see
D15 for discussion).
We then apply a global scaling of this 2-D density distribution so that the CSM and/or the ejecta has the same mass
as in the spherical model counterpart.
This is done to facilitate the comparison with spherical models and the results presented in D15.
The results from our ejecta/wind-CSM simulations are presented in Sections~\ref{sec_asymcsm} and \ref{sec_asym_ej}.

For the ejecta/disk interaction simulations, we adopt a disk density of the form
 $\rho(r,\mu) \propto 1 / r^2$ (which assumes, for simplicity, an infinite vertical scale height).
The disk extends in radius from $r_{\rm t}$ to $r_{\rm max}$ and is bounded in latitude
by the half-opening angle $\theta_{\rm D}$ above the equator (our simulation assumes top/bottom symmetry
and is limited to positive latitudes).
The disk density is then adjusted to match a specific total mass and opening angle.
For directions not intersecting the disk and locations beyond the outer edge of the ejecta ($r>r_{\rm t}$), we adopt a wind
with a mass loss rate $\dot{M}=10^{-5}$\,\msunyr\ and a constant velocity of 500\,\kms, both typical of a blue-supergiant (BSG) star.
With these parameters, the ejecta/wind interaction contributes negligibly to the radiation from the whole system, and the
wind affects negligibly the dynamics of the ejecta. The results from our ejecta/disk simulations are presented in Section~\ref{sec_disk}.

\begin{figure*}
\epsfig{file=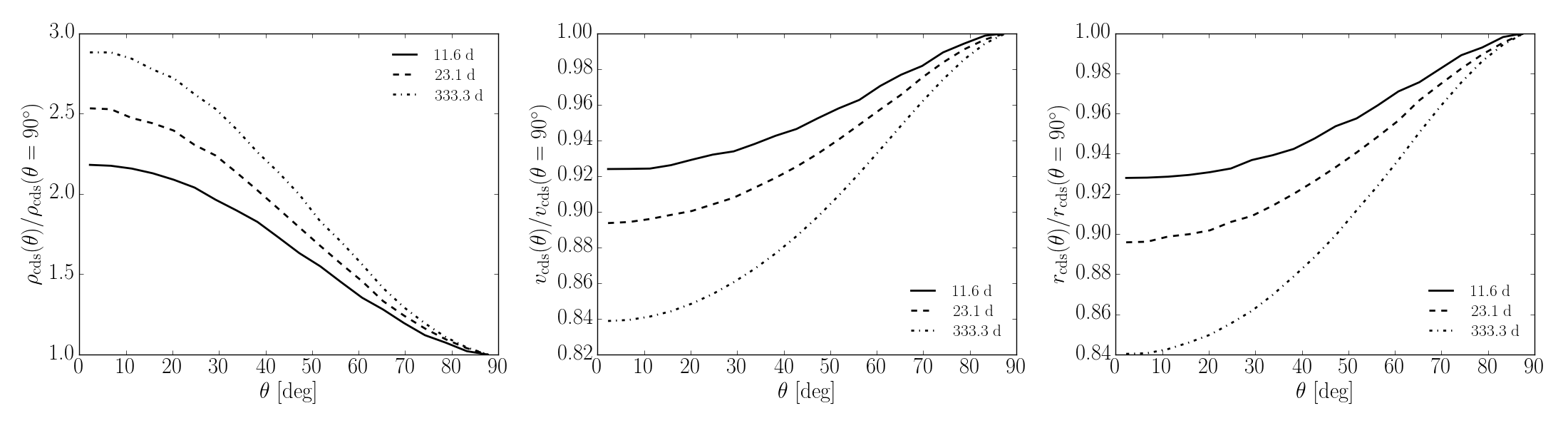,width=17.84cm}
\caption{Variation with polar angle of the CDS density (left), velocity (middle), and radius (right),
normalised to their equatorial value, and shown at 11.6, 23.1 and 333.3\,d after the onset
of interaction in model CP1.
\label{time_evol}
}
\end{figure*}

To analyse the radiative properties from our simulations, we study the radiative flux at different locations
and angles, in particular at the outer boundary to gauge the latitudinal variations.
We also study the bolometric luminosity of a model to check the conversion efficiency from kinetic
to radiative energy.
The bolometric luminosity is computed by integrating the emergent flux over the full solid angle
and is given by $4\pi \int_0^1 d\mu  r_{\rm max}^2 F_{\rm rad}(t,r_{\rm max},\mu)$, where
the radiative flux has already been integrated over energy groups.
In this paper, we do not compute the angular dependence of the observed luminosity
(this is left to future work).

Throughout this paper, we focus on signatures and features that are associated with the breaking of spherical symmetry.
For a detailed discussion of simulations for super-luminous SNe IIn arising from the interaction of a spherical  ejecta with
a spherical CSM, we refer the reader to D15 (for events like SN\,2010jl) or other works, like \citet{moriya_etal_13a}.

\section{Models with asymmetric CSM}
\label{sec_asymcsm}

\subsection{Initial conditions}
\label{init_csm}

In this section we study the interaction between a spherically symmetric ejecta and an asymmetric CSM.
The ejecta is characterised  by an initial mass of 9.8\,\msun\ and a kinetic energy of 10$^{51}$\,erg,
while the wind CSM has a total mass of 2.89\,\msun\ and expands with a constant velocity of 100\,\kms.

In model CP1, we first consider a prolate CSM density with a pole-to-equator density ratio of 3 ($A=$\,2).
We then consider a more asymmetric interaction model (CP2; $A=$\,5) and finally discuss the
differences when we switch from a prolate to an oblate CSM (model CO1; $A=-$2/3).
In all three models, we consider a CSM density distribution that varies slowly with angle ($n=$\,2).
Table\,\ref{tab_csm} gives a summary of initial properties and a few results for these simulations.

\subsection{CSM with a prolate density distribution: model CP1}

CP1 describes a SN ejecta propagating into a CSM which has a prolate density distribution and is characterised
by a pole-to-equator density ratio equal to 3. Below, we first describe some important properties
of the simulation at four representative epochs from 2.3 to 333.3\,d after the onset of interaction
(Fig.~\ref{n2_multi}). We then discuss in more detail the properties of the radiation field and
its variation with latitude, depth, and time.

\begin{figure*}
    \epsfig{file=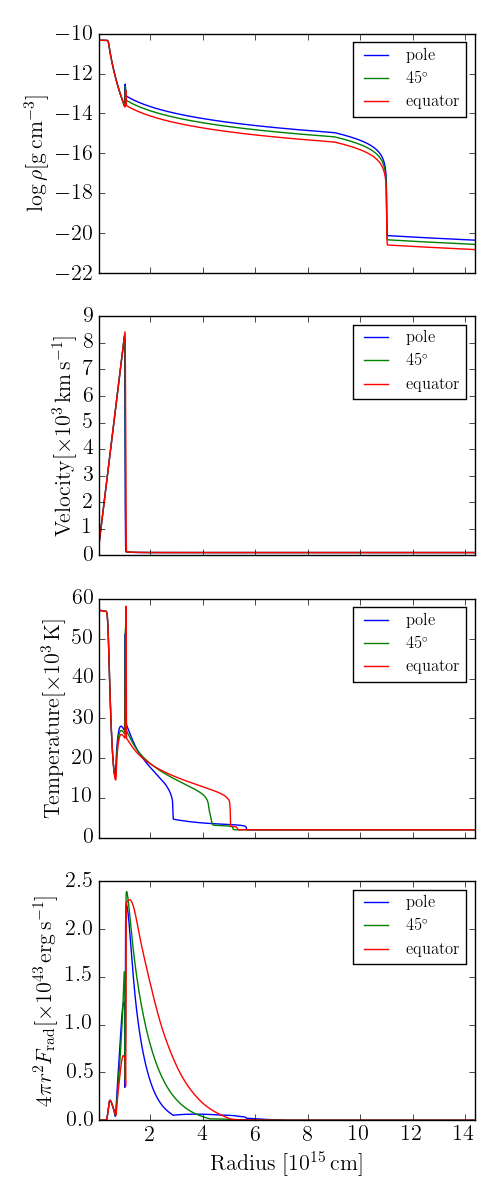,width=5.86cm}
    \epsfig{file=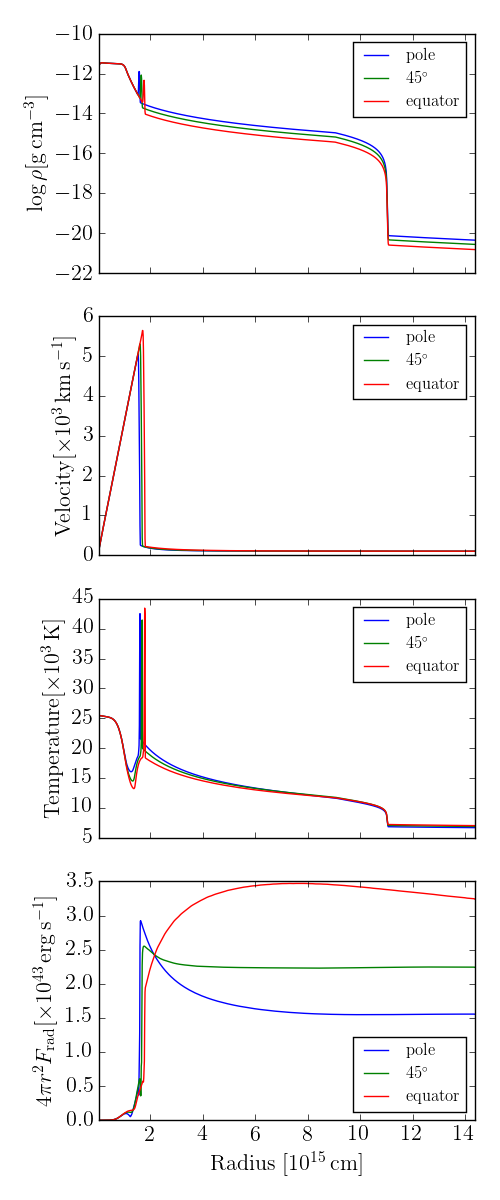,width=5.86cm}
    \epsfig{file=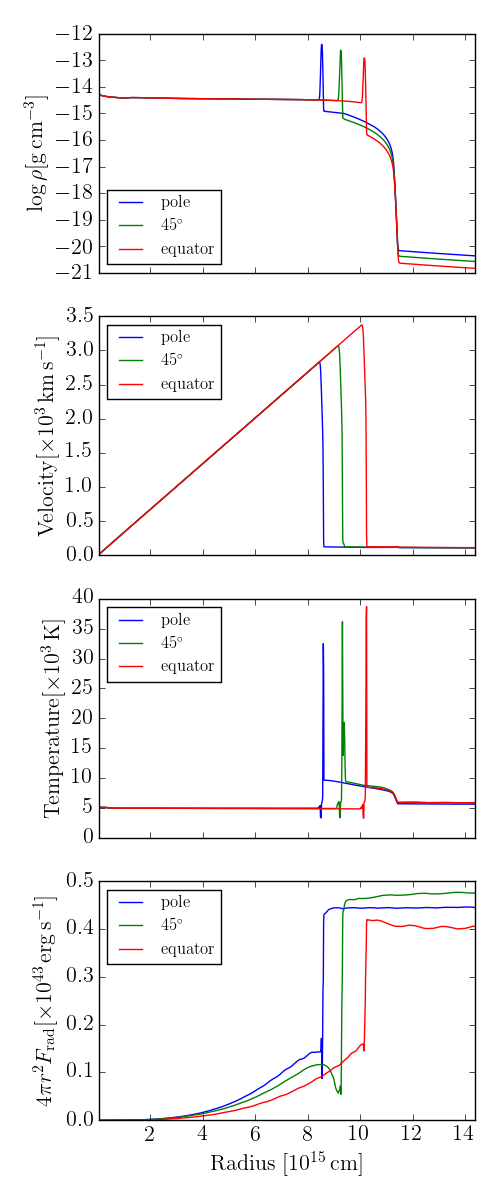,width=5.86cm}
    \caption{Radial variation of the density (top row), velocity (second row from top),
    temperature (3rd row from top), and total radiative flux (bottom row)
    along the polar, 45$^\circ$, and equatorial directions for model CP1
    at 2.3 (left), 23.1 (middle) and 333.3\,d (right) after the onset of interaction.
    The total radiative flux is scaled by $4 \pi r^2$ for easier comparison to a luminosity and to cancel the effect of spherical dilution.
    [See text for discussion.]
    \label{fig_CP1_cuts}
    }
\end{figure*}

The top-left panel of Fig.~\ref{n2_multi} is a montage of four quadrants showing the 2-D distribution at 2.3\,d
of the temperature, density, radiative energy, and radiative flux (scaled by $r^2$ to cancel the effect of spherical dilution).
We overlay the location of the photosphere (solid line, bottom-left panel), which we define for simplicity as the location
where the inward-integrated (from $r_{\rm max}$) radial optical depth due to electron scattering is equal to 2/3.
At the start of the interaction, the CSM is cold and recombined (i.e., neutral) so that the (electron-scattering) photosphere
lies within the (unshocked) ejecta (at $r<r_{\rm t}$). By 2.3\,d, the radiation from the shock propagates through the CSM,
raising its temperature and its opacity.
This ``precursor" radiation progresses faster along the lower density equatorial regions (bottom left quadrant),
causing the temperature to rise faster there, causing the photosphere to appear oblate, out of phase with the density distribution,
which is prolate. The morphology of the photosphere at 2.3\,d is therefore controlled primarily by the change in ionisation (which
can cause a change in opacity by orders of magnitude), rather than the asymmetry in density (which can cause at most a
change of factor of 3 in opacity for this model CP1).
The radiative precursor first reaches the outer boundary along the equator,  at $t=$\,5.8\,d, and at increasingly later times
for higher latitudes (at $t=$\,9.3\,d along the pole).

\begin{figure*}
\epsfig{file=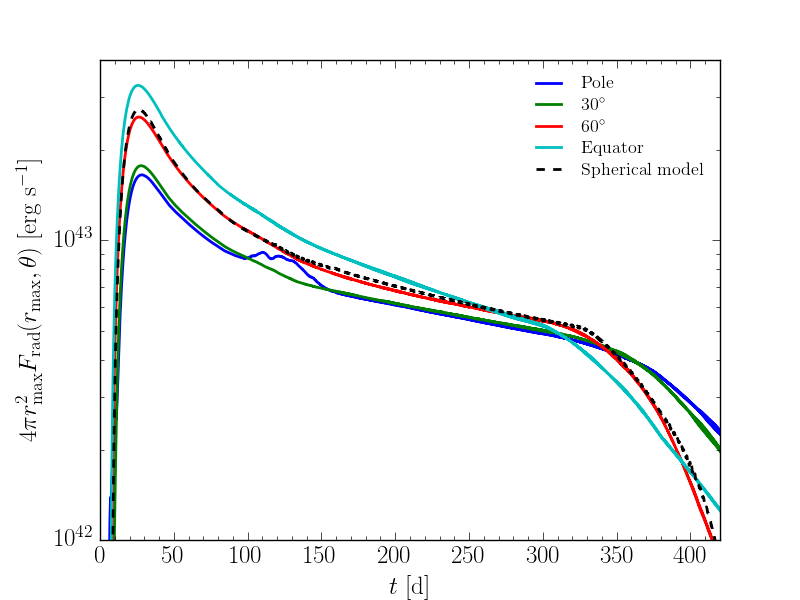,width=8.8cm}
\epsfig{file=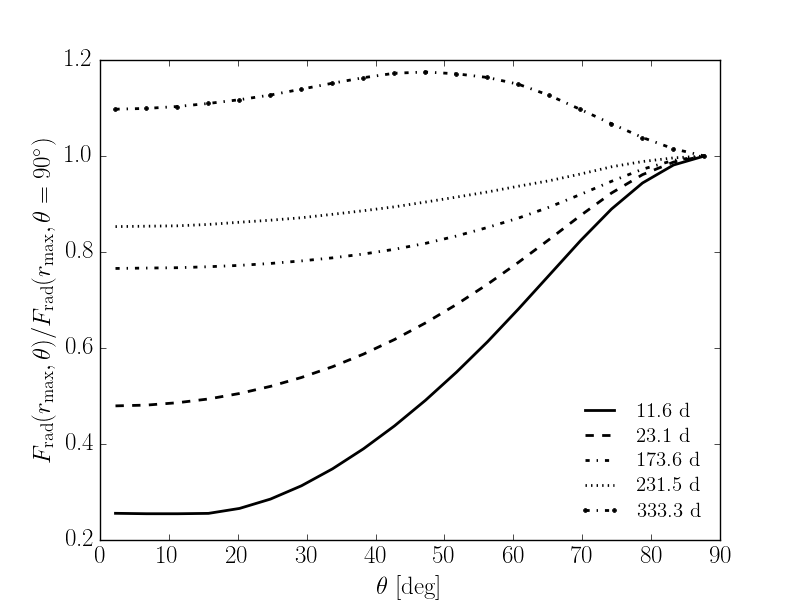,width=8.8cm}
  \caption{
 Left: Evolution of the luminosity-like quantity $4\pi r_{\rm max} ^2 F_{\rm rad}(r_{\rm max}, \theta)$
 along $\theta=$\,0 (pole), 30, 60, and 90\,deg (equator) for model CP1, together with the bolometric luminosity
 for the spherical counterpart (dashed line). Notice how the flux is initially greater along the equator, where the radial
 optical depth is lower, and eventually becomes stronger along the poles, where the ejecta deceleration is stronger.
Right: Variation with latitude and at selected times of the emergent radiative flux $F_{\rm rad}(r_{\rm max}, \theta)$
(normalised to its equatorial value).
\label{lc_n2}
}
\end{figure*}

The top-right panel of Fig.~\ref{n2_multi} shows  the properties of model CP1 at 23.1\,d, which corresponds to
the epoch of flux maximum along the equator.
The radiation from the shock has now completely filled the grid and the temperature has risen everywhere
above $\sim$10\,000\,K. Hydrogen, the main electron donor in the CSM, is now ionised
and the CSM optical depth is large. The morphology of the photosphere is now prolate, just like the density distribution,
because there is no ionisation bias between different latitudes. The photosphere is located far above the shock along
all directions, and about 50\% further out along the pole than along the equator.
Similarly, the radial electron-scattering optical depth at the shock is about 10 along the pole and a third of that along the equator
(because of the imposed density contrast).
The first striking property for this asymmetric CSM is the $\sim$\,50\% larger flux at $r_{\rm max}$ along the equator
compared to the pole, while at the shock the polar flux is actually greater than the equatorial flux (bottom-left panel).
In our simulation, the luminosity at the shock (and the radiative energy immediately above the shock) is greater
where the ejecta deceleration by the CSM is greater, hence at higher latitudes.
Photons originally released at the shock in the X-ray and UV range are absorbed and re-emitted as lower energy
(UV and optical) photons.
These photons then diffuse through the CSM, taking the path of least resistance, and preferentially emerge
 along the equatorial regions. The distribution of the flux is thus out-of-phase with the distribution of mass (or scatterers)
 and the distribution of the radiation-energy density at the shock.
This effect is pronounced early on, even though the CDS is only weakly asymmetric, showing an elongation towards
the lower-density equatorial regions where the ejecta deceleration is smaller ($(r_{\rm p}/r_{\rm e})_{\rm CDS}=$\,0.9).
The deformation of the CDS takes longer because it takes time to plow through a sizeable chunk of this very extended CSM.

The bottom left panel of Fig.~\ref{n2_multi} shows the results for model CP1 at $t=$\,138.9\,d.
At that time, the CDS is about to overtake the photosphere along the equator --- it will do so at $\sim$\,250\,d
along the polar direction.
Because the CSM is nearly optically thin above the CDS along the equator, radiation energy is inefficiently stored
there (top-left quadrant) and the flux contrast between polar and equatorial directions is much reduced.
At this time, the photosphere has the same prolate morphology as at 23.1\,d, but the CDS exhibits an oblate
morphology with a pole-to-equator radius ratio of $\approx$\,0.85.

The bottom right panel of Fig.~\ref{n2_multi} shows the results for model CP1 at $t=$\,333.3\,d.
Both the CSM and the CDS are optically thin along the equator, and the photosphere now progressively
recedes into the unshocked ejecta along that direction. The transition to optical thinness occurs with
a greater delay along higher latitudes (at 380\,d along the pole).
There is now no strong optical-depth effect acting on the shock luminosity.\footnote{To be precise, X-ray and UV photons
are still absorbed locally because of their short mean free path. This radiative energy is, however, drained by optical photons,
which escape freely --- see D15 and their Fig.~7 for details.}
The situation is essentially steady state (no retardation due to diffusion) so that the emergent
luminosity is equal to the shock luminosity and is now greater at higher latitudes (probably because of residual optical
depth effects, the flux reaches a maximum at mid-latitudes rather than along the pole).

The interaction of a spherical SN ejecta  with an asymmetric CSM therefore exhibits interesting new features
absent in 1-D models.
While the photosphere location settles far out in the CSM within a week of the onset of the interaction, the
CDS density, velocity, and radius continuously evolve as more material is swept up by the shock (Fig.~\ref{time_evol}).
The CDS density varies in proportion to the swept up mass. At the end of the simulation, the pole-to-equator density ratio
of the CDS is about 3, which is equal to the pole-to-equator density ratio (at a given $r$) adopted initially in model CP1.
The equality is not exact because the velocity of the CDS is slower along higher latitudes.
In our initial setup, the ejecta/CSM interface is spherical, and the CDS that quickly forms becomes increasingly
oblate, with $(r_{\rm p}/r_{\rm e})_{\rm CDS}\sim$\,0.89 at bolometric maximum and $\sim$\,0.84 at 333.3\,d
(Fig.~\ref{time_evol}). Consequently, the ratio $(r_{\rm Phot}/r_{\rm CDS})(\theta)$ evolves
and the photons emitted around $r_{\rm CDS}(\theta)$ are subject to an angle and time-dependent optical depth.
In other words, the cocoon of optically-thick material above the CDS has a complex and evolving shape (Fig.~\ref{n2_multi}).
Similarly, the shape of the electron-scattering photosphere is clearly prolate early on but switches to oblate when the CDS
overtakes it at late times. These properties are important for understanding the polarisation signatures of SNe IIn like SN\,2010jl
(\citealt{Patat_2011}; see below).

In Fig. \ref{fig_CP1_cuts} we show the radial variation of the velocity, density, temperature,
and total radiative flux (scaled by $4 \pi r^{2}$ for easier comparison to a luminosity
and to cancel the effect of spherical dilution) for different times.
At 2.3\,d, the radiation from the interaction is slowly filling up the grid.
The CSM is cold and optically-thin to optical photons, facilitating the propagation of the radiation along the lower-density
equatorial directions. This epoch corresponds to the ``radiative-precursor" phase.
At 23.1\,d, the CSM is optically thick to all photons, irrespective of their energy. In the absence of
optical depth effects, and provided that the shock luminosity varies slowly over the diffusion time,
the quantity $r^2 F_{\rm rad}$ would just be constant from the shock until the outer boundary.
Instead, $r^2 F_{\rm rad}$ decreases along the polar direction and increases along the equatorial direction, the
latter at the expense of the former. The optically thick CSM redistributes the flux in angle, and reverses
the pole-to-equator flux ratio between the shock and the outer boundary.
The flux is strongly redistributed from the polar direction, where the deceleration is stronger, towards the equatorial
regions, where the photon mean free path is larger.
This redistribution is caused by the latitudinal dependence of the radial optical depth and confirms
the previous results of \citet{Dessart_2011}, which were based on 2-D radiative-transfer simulations (without dynamics).
At 333.3\,d, the CSM is optically thin and the shock luminosity is essentially constant with radius, but stronger
along higher latitudes where the CSM is denser.

Fig.~\ref{lc_n2} illustrates the latitudinal dependence of the radiative flux at the outer boundary,
at all times and for a few polar angles (left panel), as well as at selected times and for all polar angles (right panel).
The flux along $\theta=$\,60\,deg closely matches its counterpart from the equivalent spherical model
(dashed line). The flux maximum is 28\% greater along the equator and 36\% lower along the pole
compared to the spherical model.
Bolometric maximum occurs at 26.4\,d after the onset of  interaction --- the flux maximum along the
equator occurs 3-4\,d earlier than along the polar direction. This small delay is somewhat surprising since the ``radial'' diffusion
time is about 3 times longer along the pole than along the equator. As discussed above and shown in Fig.~\ref{fig_CP1_cuts},
photons leak from the denser polar regions and escape through the equatorial regions. Using the radial optical
depth only gives a ``radial" diffusion time, which is different from the effective diffusion time through the asymmetric CSM.
At the time of maximum, the emergent flux is about 40\% greater along the equator than along the pole,
and the offset decreases steadily towards higher latitudes.
The pole/equator offset also decreases continuously with time until the CSM turns optically thin,
when photon redistribution in angle ceases.
Consequently, at late times, the flux is greater in the polar regions, i.e., where the shock luminosity is greater.
The contrast is also exacerbated by the fact that the shock progresses more slowly at higher latitudes,
so that it interacts with denser material for longer.
The breaks at $\gtrsim$\,300\,d (equator) and $\sim$\,350\,d (pole) correspond to the time
when the shock leaves the dense
part of the CSM along those directions.\footnote{At high latitudes, the flux exhibits a bump at 100-150\,d.
This bump makes no sense physically. We could not identify the origin of the problem, but we find that
by increasing the number of radial points to 2400, the bump disappears.
This higher resolution simulation gets stuck into small timesteps at late times so we show instead
the simulation that uses 1200 radial points.}
So, despite the prolate CSM density distribution, the emergent flux is initially greater along the equatorial
direction (when the CSM is optically thick) but is progressively biased in favour of higher latitudes
later on (the flux is actually maximum along mid-latitudes at $\sim$\,350\,d).

Model CP1 was evolved for 500\,d. At the end of the simulation, the time-integrated radiative flux
through the outer boundary is $0.3\times10^{51}$\,erg, which corresponds to 30\% of the initial ejecta
kinetic energy. The spherical model counterpart yields the same conversion efficiency, probably because
the CSM (spherical or not) has the same total mass in both simulations and the magnitude of the asymmetry
in model CP1 is moderate.

\begin{table*}
\caption{Same as Table \ref{tab_csm}, but now for simulations of the interaction between an asymmetric
ejecta and a symmetric CSM.
\label{tab_asymej}}
\begin{tabular}{l@{\hspace{7mm}}c@{\hspace{4mm}}c@{\hspace{4mm}}
c@{\hspace{4mm}}c@{\hspace{4mm}}c@{\hspace{4mm}}c@{\hspace{4mm}}
c@{\hspace{4mm}}c@{\hspace{4mm}}c@{\hspace{4mm}}c@{\hspace{4mm}}}
\hline
Model
&  Ejecta
& $A$
& $n$
& $(\rho_{\rm p}/\rho_{\rm e})_{t=0}$
& $t_{\rm peak}$ [d]
& $(r_{\rm p}/r_{\rm e})_{\rm Phot}$
& $(r_{\rm p}/r_{\rm e})_{\rm CDS}$
& $(\upsilon_{\rm p}/\upsilon_{\rm e})_{\rm CDS}$
& $(\rho_{\rm p}/\rho_{\rm e})_{\rm CDS}$
&$(F_{\rm p}/F_{\rm e})_{r_{\rm max}}$ \\
\hline
EP1           &    prolate         &    2     &   2   &   3       &  26.2 & 1.0   &  1.05  & 1.11 & 1.20 &  1.02     \\
EO1           &    oblate          &  -5/6  &   2   &  1/6      &  26.1 & 1.0  &   0.76  & 0.85 & 0.80 & 0.98 \\
EP2           &    prolate         &   199  &  50   &   200   &  25.1 & 1.0   &  1.56   & 1.64 & 2.37 & 1.06 \\
\hline
\end{tabular}
\end{table*}

\begin{figure*}
\epsfig{file=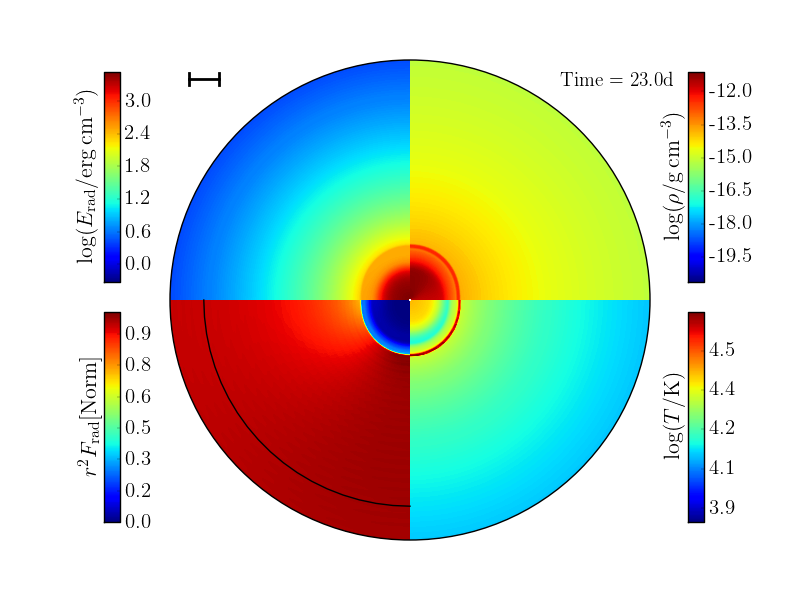,width=8.8cm}
\epsfig{file=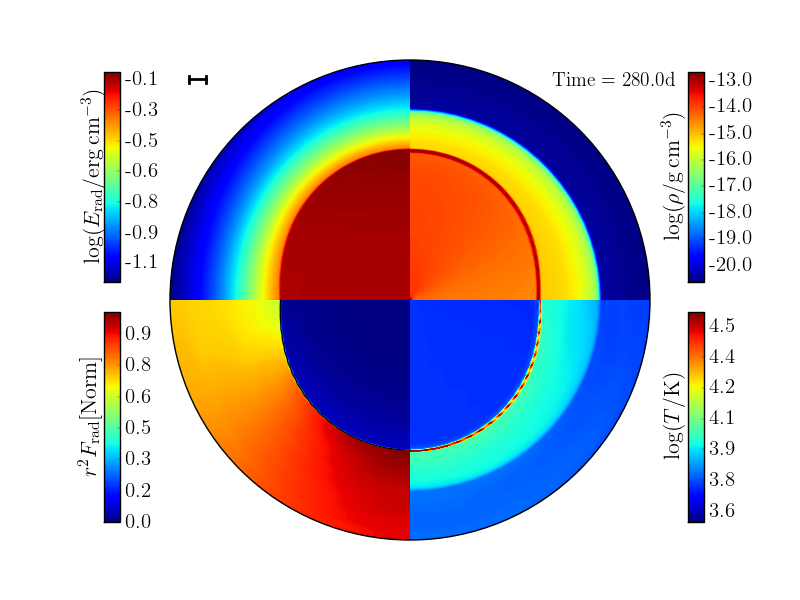,width=8.8cm}
\caption{
Same as Fig.~\ref{n2_multi}, but now for model EP1 at  23.0 (left) and 280.0\,d (right;
the photosphere resides within the CDS along all latitudes).
The length of the thick bar in each panel corresponds to a physical scale of 10$^{15}$\,cm.
In contrast to model CP1, the spherically-symmetric CSM tends to damp the latitudinal variation
of the flux at the shock at early times.
As the CSM optical depth drops above the CDS, the emergent flux becomes more latitude dependent.
\label{m1_20d}
}
\end{figure*}

\subsection{Other asymmetric CSM configurations}
\label{sect_asym_csm_other}

   The results presented above for model CP1 are relevant for a wide variety of simulations
   we performed for a spherical ejecta interacting with an asymmetric CSM, provided
   the imposed asymmetry is moderate --- differences in the results are at the quantitative rather
   than the qualitative level.

   Keeping the same prolate CSM density distribution but enhancing the pole-to-equator density
   ratio exacerbates the latitudinal dependence of the radiation and the gas.
   Model CP2 ($A=$\,5) yields a CDS that is denser and slower along the pole than in model CP1.
   The CDS morphology becomes more oblate, with
   $(r_{\rm p}/r_{\rm e})_{\rm CDS}=$\,0.75 at 500\,d compared to 0.82 in model CP1.
   While the CSM is optically thick, the photosphere is more prolate and the flux redistribution
   towards the equatorial regions is even greater than in model CP1.
   At $t=$\,23.0\,d, the pole-to-equator ratio of the photospheric radius is 2.27 for model CP2
   compared to 1.61 for model CP1,
   and the pole-to-equator ratio of the emergent radiative flux is 0.30 for model CP2
   compared to 0.47 for model CP1.

   If we reverse the asymmetry of the CSM from prolate to oblate (model CO1),
   the asymmetry in the radiative flux, the CDS velocity and density, and the ratios
   $(r_{\rm p}/r_{\rm e})_{\rm CDS}$, $(r_{\rm p}/r_{\rm e})_{\rm Phot}$ are reversed.
   For example, the shock now propagates faster along the poles than the equator and
   the lower CSM density at higher latitudes biases the escape of photons towards the pole
   (as long as the overlying CSM remains optically thick).
   A summary of the main results for the asymmetric CSM simulations is given in Table \ref{tab_csm}.

\section{Models with asymmetric ejecta}
\label{sec_asym_ej}

\subsection{Initial conditions}
\label{init_ej}

We now turn to interactions involving an asymmetric ejecta and a symmetric CSM.
In model EP1 (EO1), the ejecta has a prolate (oblate) density distribution with a pole-to-equator density ratio of 3 (1/6).
We also consider model EP2 where the asymmetry is more pronounced by being more confined to the pole (by increasing $n$
from 2 to 50) and by having a greater pole-to-equator density contrast (by increasing $A$ from 2 to 199) --- this
model is done to mimic a bipolar explosion ramming into a spherical CSM.\footnote{Because the inner shell is an
ejecta in homologous expansion, the ejecta velocity at a given radius is the same for all latitudes. Imposing an asymmetry
on the velocity field to mimic an asymmetric explosion would thus require a non-spherical interface between ejecta and
CSM, which we choose not to use in our setup. Instead, we start from a spherical ejecta/CSM interface and impose a latitudinal
dependence of the ejecta density distribution.}

\subsection{Moderate-asymmetric ejecta}

Model EP1 is the counterpart of model CP1, but now the asymmetry is tied to the ejecta rather than the CSM.
This change has a dramatic impact on the behaviour of the radiation field.

Figure~\ref{m1_20d} shows the same montage as Fig.~\ref{n2_multi} but now for model EP1.
Compared to model CP1, the latitudinal dependence of all quantities is much weaker.
At 23.0\,d after the onset of interaction (left panel), the radiative flux at $r_{\rm max}$ varies
by less than 2\% with latitude, and at this early stage, the CDS has too recently formed to show
a sizeable asphericity ($(r_{\rm p}/r_{\rm e})_{\rm CDS}=$\,1.05).

At 280.0\,d (right panel of Fig.~\ref{m1_20d}),
the CDS has a clear prolate morphology ($(r_{\rm p}/r_{\rm e})_{\rm CDS}=$\,1.10
and $(v_{\rm p}/v_{\rm e})_{\rm CDS}= $\,1.16) --- the denser ejecta regions
have more inertia and are decelerated less by the CSM.
However, $(\rho_{\rm p}/\rho_{\rm e})_{\rm CDS}\sim$\,1.0 since the swept up matter is uniformly distributed.
The pole-to-equator flux ratio at the outer boundary is 1.38, and has been growing since the start of the interaction.
In this simulation, although the shock luminosity varies with latitude, the spherically symmetric CSM
is optically thick and damps this variation --- as long as the CSM is optically thick, the photosphere is
located at $6\times10^{16}$\,cm along all latitudes.
This damping is maximum at the onset of the interaction.
As the CDS progresses outwards, the overlying CSM becomes less optically thick and the redistribution/diffusion
of the radiation weakens, letting a residual angle dependence of the flux survive at the outer boundary.
The pole-to-equator flux ratio at $r_{\rm max}$ is sizeable once the photosphere resides within the CDS.

\begin{figure}
\epsfig{file=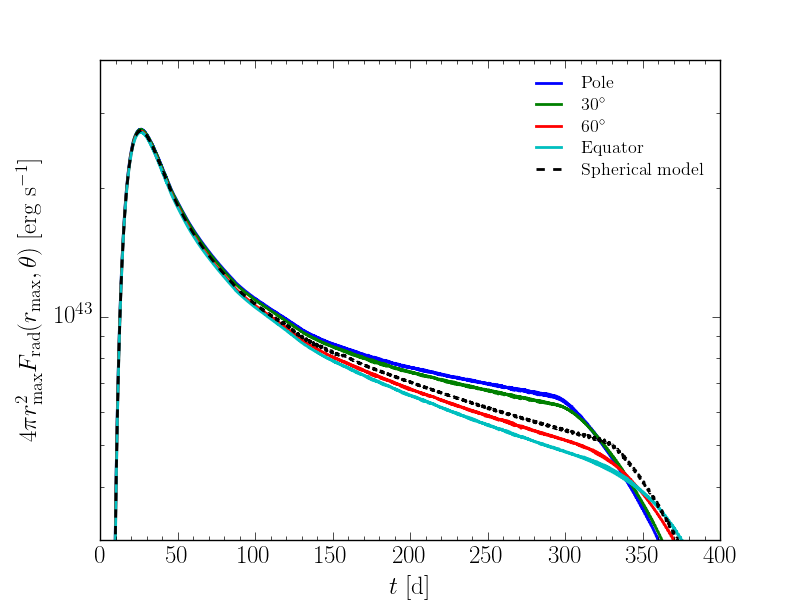,width=8.8cm}
\caption{
Evolution of the luminosity-like quantity $4\pi r_{\rm max} ^2 F_{\rm rad}(r_{\rm max}, \theta)$
 along $\theta=$\,0 (pole), 30, 60, and 90\,deg (equator) for model EP1, together with the bolometric luminosity
 for the spherical counterpart (dashed line).
 The emergent flux shows no latitude dependence up until $\sim$150\,d. After that, the CSM optical
 depth above the CDS decreases and the latitudinal dependence of the shock luminosity, although
 possibly damped by the spherically-symmetric CSM, survives at the outer boundary --- this holds
 in particular once the CDS overtakes the photosphere at 160\,d.
\label{lc_m1}}
\end{figure}

This latitudinal variation of the emergent flux is shown in Fig.~\ref{lc_m1}.
Despite the asphericity of the interaction, the emergent radiation is isotropic at bolometric maximum.
A latitudinal dependence appears after 150\,d, when the CDS comes close to the photosphere in the CSM,
and eventually overtakes it (at 160\,d along the pole and 25\,d later along the equator).
The break at $\sim$\,300\,d for all light curves correspond to the time when the shock leaves
the dense part of the CSM.

We have also performed a simulation for an oblate ejecta interacting with a symmetric CSM (model EO1).
The results follow the same principles as those presented for the comparison between
models CP1 and CO1 (Section~\ref{sect_asym_csm_other}; see also Table~\ref{tab_asymej}).

\begin{figure*}
  \epsfig{file=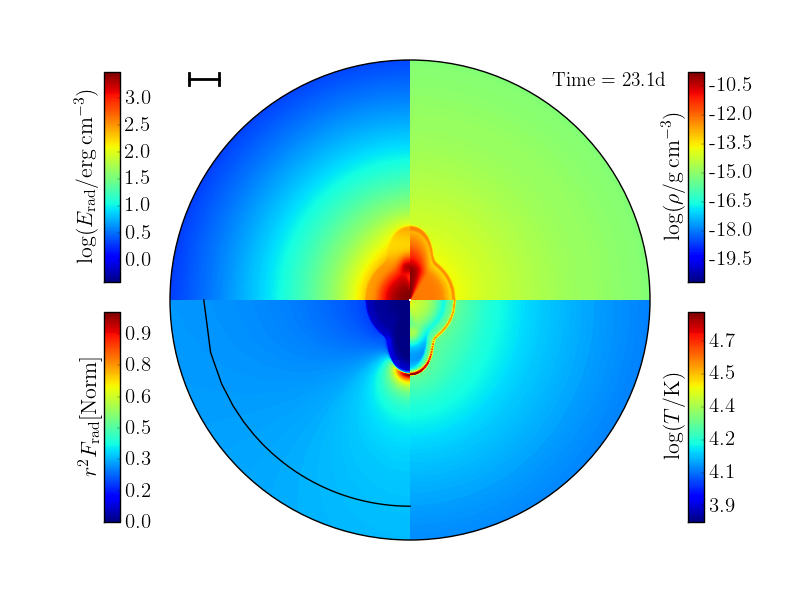,width=8.8cm}
  \epsfig{file=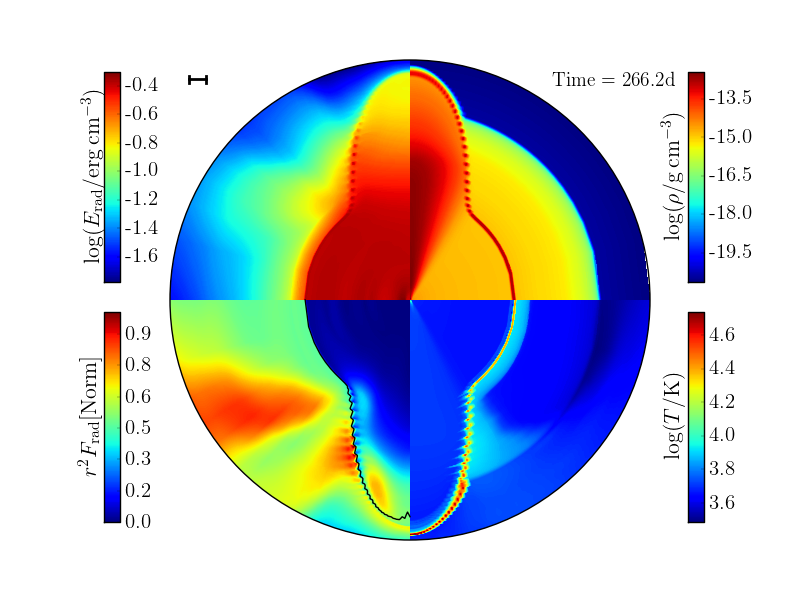,width=8.8cm}
  \caption{Same as Fig.~\ref{n2_multi}, but now for model EP2 at 23.1 (left) and 266.2\,d (right)
  after the onset of interaction.
 The length of the thick bar in each panel corresponds to a physical scale of 10$^{15}$\,cm.
  Notice how the angular variation of the flux at the shock is greatly attenuated early on by the optically-thick
  spherically-symmetric CSM.
  \label{jet_sn}
  }
\end{figure*}

\begin{figure*}
  \epsfig{file=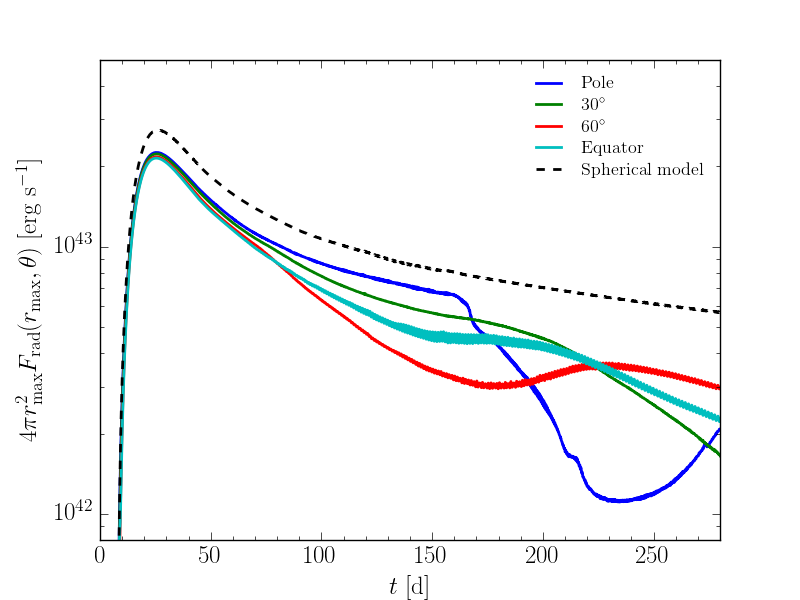,width=8.8cm}
  \epsfig{file=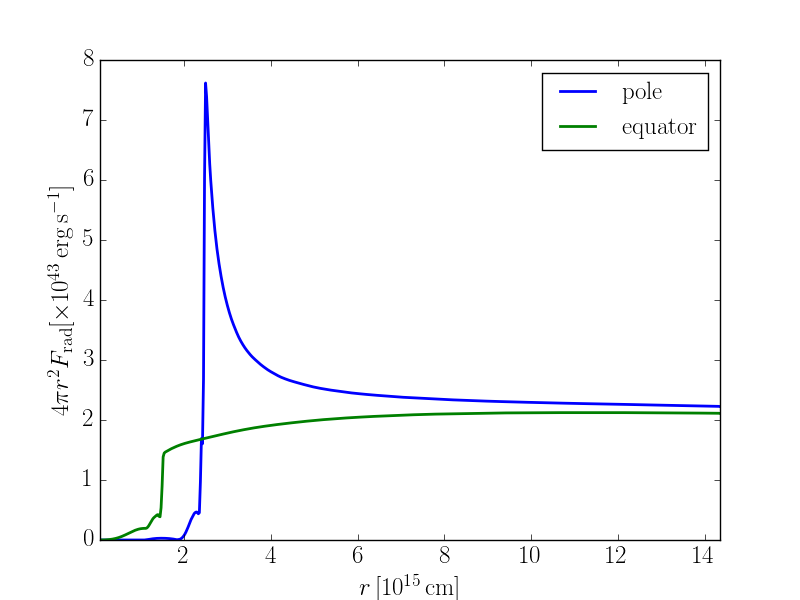,width=8.8cm}
  \caption{
  Left: Evolution of the luminosity-like quantity $4\pi r_{\rm max} ^2 F_{\rm rad}(r_{\rm max}, \theta)$
 along $\theta=$\,0 (pole), 30, 60, and 90\,deg (equator) for model EP2 (the dashed line corresponds to the
 spherical model counterpart).
 The pole-to-equator flux ratio is limited to a few percent at maximum but grows to reach a factor
 of two at 150\,d. This ratio drops below unity when the CDS leaves the dense part of the CSM along
 the poles, while interaction persists for longer along lower latitudes.
  The rebrightening at 230\,d along the poles occurs when that material becomes optically thin in that direction,
  allowing the escape of stored radiation energy at depth.
  Right:  Radial variation of the total radiation flux scaled by
  $4\pi r^{2}$ at $t=$\,23.1\,d. Notice the strong redistribution of the flux by the optically thick CSM,
  causing a negligible flux contrast with latitude at $r_{\rm max}$.
  \label{jet_lc}
  }
\end{figure*}

\begin{figure*}
\epsfig{file=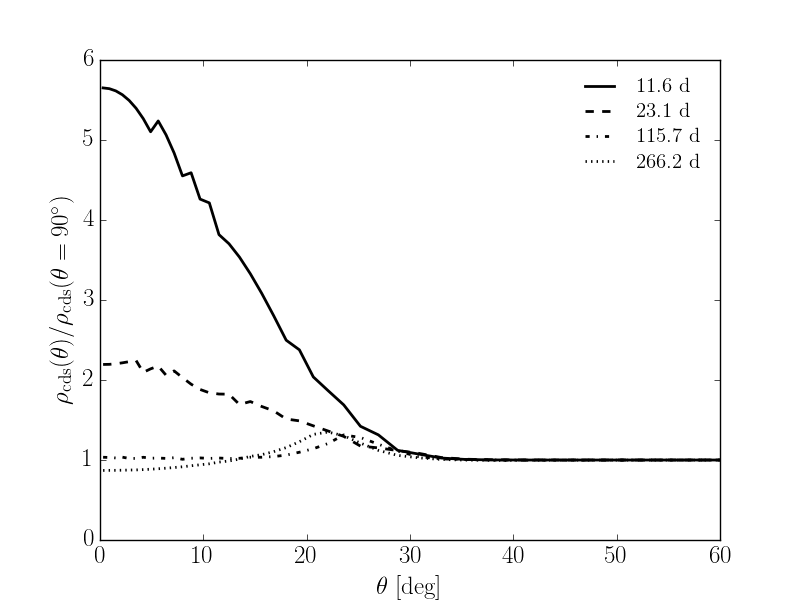,width=8.8cm}
\epsfig{file=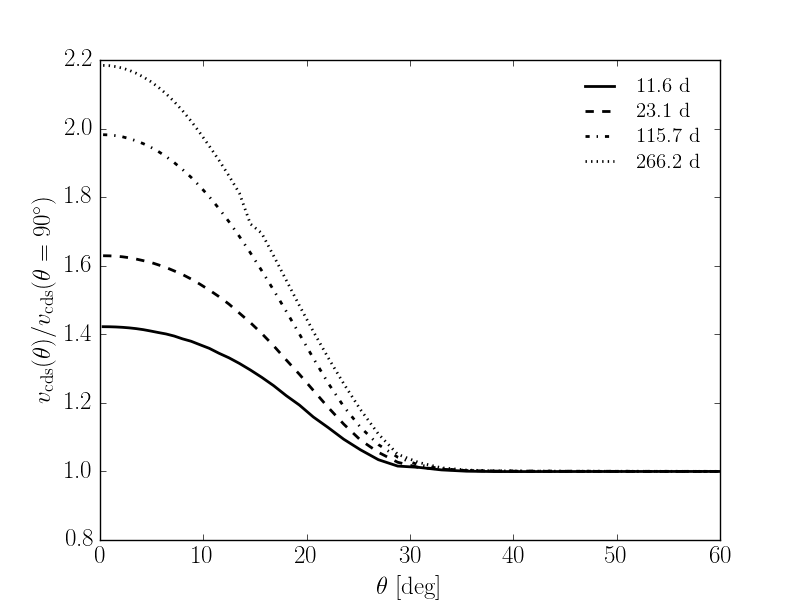,width=8.8cm}
\caption{
Variation with polar angle of the CDS density (left) and velocity (right)
normalised to their equatorial value, and shown at 11.6, 23.1 and 115.7, and 266.2\,d after the onset
of interaction in model EP2.
\label{jet_cds_comp}
}
\end{figure*}

\subsection{Highly-asymmetric ejecta}

We now consider the interaction of a highly-asymmetric ejecta ($A=$\,199 and $n=$\,50)
with a spherically symmetric CSM (model EP2; Table~\ref{tab_asymej}).
We show two snapshots of the \heracles\ simulation at 23.1 and 266.2\,d after the onset of
interaction in Fig.~\ref{jet_sn}, together with the emergent flux as a function of latitude
in Fig.~\ref{jet_lc}.

At 23.1\,d (left panel of Fig.~\ref{jet_sn}), the radiation from the shock has already raised
the temperature, ionisation, and opacity in the CSM so that it is optically thick.
The strong asymmetry adopted for the ejecta density produces a strongly asymmetric (bipolar)
explosion, with $(r_{\rm p}/r_{\rm e})_{\rm CDS}=$\,1.6 and $(v_{\rm p}/v_{\rm e})_{\rm CDS}=$\,1.6
at that time (Fig. \ref{jet_cds_comp}).
The interaction is much stronger at higher latitudes so that the shock luminosity is maximum along
the pole --- we have $(F_{\rm p}/F_{\rm e})_{\rm CDS}=$\,5.3 (right panel of Fig.~\ref{jet_lc}).
However, the large optical depth of the spherically-symmetric CSM damps this variation
(which is much larger than in model EP1).
At 23.1\,d after the onset of the interaction (left panel of Fig.~\ref{jet_lc}),
the flux at the outer boundary is constant with latitude within a few percent.

At 266.2\,d (right panel of Fig.~\ref{jet_sn}), the interaction has evolved into a very complex
aspherical structure. The shock has entirely crossed the dense part of the CSM along the pole
(the CDS leaves the grid along the pole at 280.0\,d and at that time we have
$(r_{\rm p}/r_{\rm e})_{\rm CDS}=$\,2.3)
while it is only halfway through the dense CSM along the equator.
The shock deceleration along the pole is now negligible, while it is still significant along the equator.
The CSM is optically thick along the equator but optically thin along the poles.
The flux varies strongly with latitudes and is now lower along the poles.
As time progresses, the dominant flux contribution comes from lower and lower latitudes,
where the interaction persists for longer.
There is a rise in the polar flux at $\gtrsim$\,240\,d because the polar direction becomes
optically thin, allowing the radiative energy stored at depth to escape (the equatorial regions
are optically thick and prevent this escape).

Compared to the spherical counterpart, model EP2 has at all times a lower flux, whatever the latitude
considered (Fig. \ref{jet_lc}, left panel), and its conversion efficiency is 13\% lower.
The main reason for this is the weaker deceleration of the ejecta at high latitudes. The CSM is not dense
and extended enough to slow the material down so that a larger fraction of the ejecta kinetic energy remains
untapped.

In model EP2, the large optical depth of the ejecta should lead to the formation of
narrow lines with broad electron scattering wings early on. As the fast moving inner ejecta
progresses along the pole towards the outer CSM, broad lines should also appear.
Along certain viewing angles, this configuration may produce a triple-peak H$\alpha$ profile and share some of
the properties of the type IIn SN\,2010jp \citep{smith_10jp_12}.

\section{Symmetric SN ejecta - Disk CSM}
\label{sec_disk}

\subsection{Initial conditions}
\label{init_setupcsm}

   We now consider the interaction between a type II SN ejecta (with the same properties as used for model CP1)
and a relic disk extending outwards  from $r_{\rm t}=$\,10$^{15}$\,cm.
Outside the disk and beyond the ejecta outer edge at $r_{\rm t}$, we fill the grid with a 10$^{-5}$\,\msunyr\ BSG wind
 (see Section~\ref{sect_radhydro} for the details on the numerical setup).

   Model D1 has a disk with a half opening angle  $\theta_{\rm D}$ of 5\,deg and a total mass equal to 1.5\,\msun.
Model D2 has a disk with a half opening angle of 5\,deg but a total mass equal to 5\,\msun.
Model D3 has a disk with a half opening angle of 10\,deg and the same total mass of 1.5\,\msun\ as in model D1.
In each case, we apply a global scaling to the disk density in order to match the desired disk mass.
We obviously do not consider all possible configurations. For example, we leave aside the possibility that the disk
starts at a smaller or larger radius than 10$^{15}$\,cm. With the present model set, we capture some
of the salient features of ejecta/disk interactions.

To quantify the contribution of the ejecta-disk interaction in our simulations, we run an additional (spherically-symmetric)
model without the disk (model SNW).

\subsection{Results for the ejecta/disk model D1}
\label{results_disk}

\begin{figure*}
   \epsfig{file=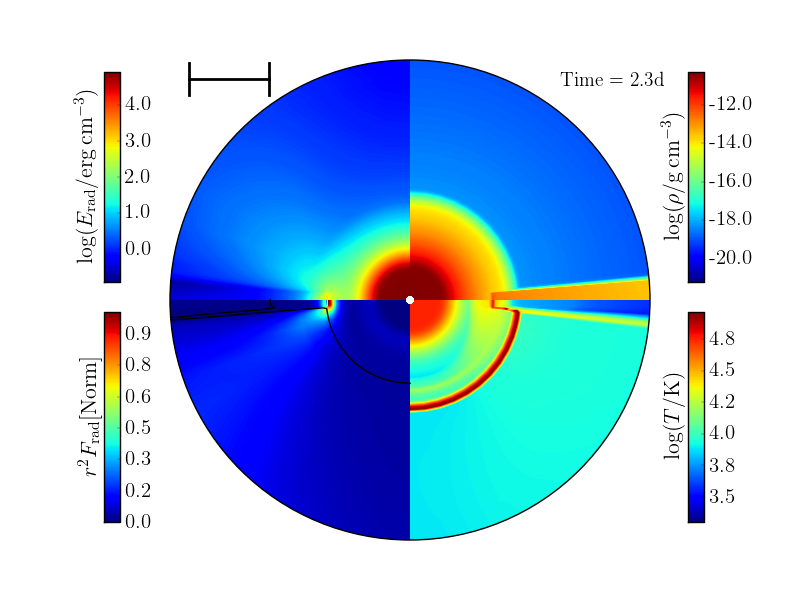,width=8.8cm}
   \epsfig{file=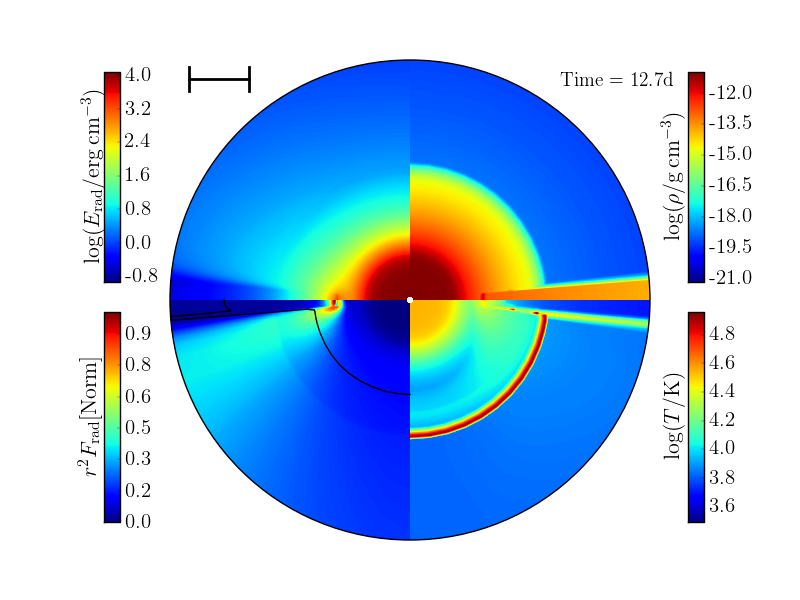,width=8.8cm}
   \epsfig{file=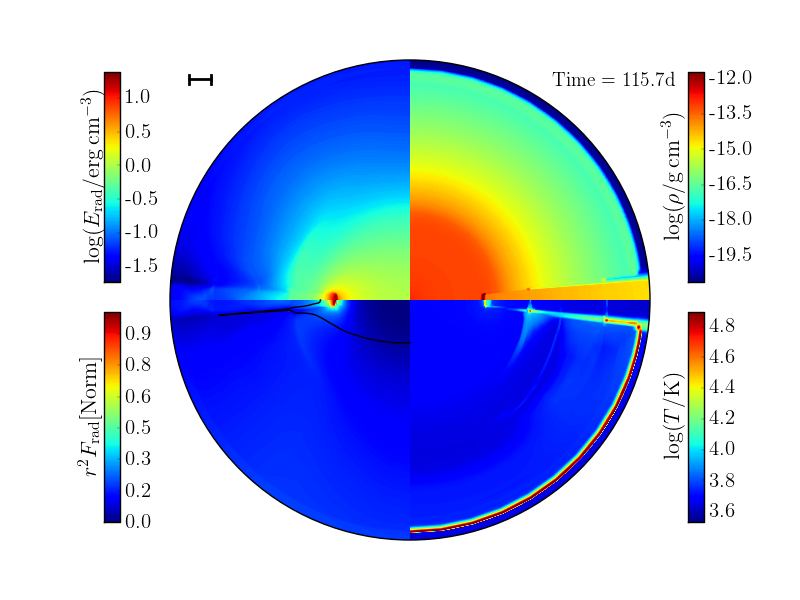,width=8.8cm}
   \epsfig{file=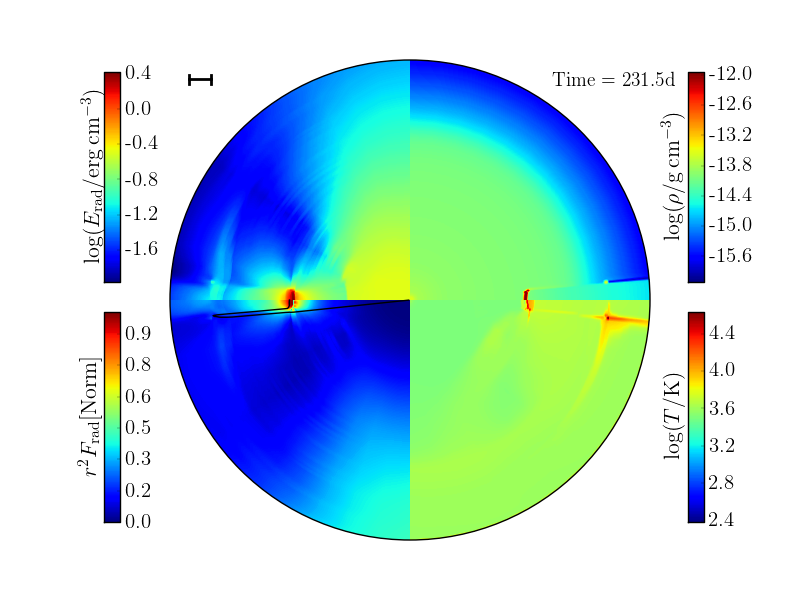,width=8.8cm}
   \caption{
  Top left:
      Dynamical and radiative properties of the ejecta/disk model D1 at 2.3\,d after the onset of interaction.
   The four quadrants show the 2-D properties of the density, the temperature, the radiative flux (multiplied by $r^{2}$ and
   normalised; we overlay the electron-scattering photosphere as a solid line)
   and the radiative energy.
  Here and in other panels, the length of the thick bar corresponds to a physical scale of 10$^{15}$\,cm.
    At this early time, the stored radiation leaking from the ejecta is filling up the grid while the radiation
   injected at the shock diffuses through the optically-thick material that engulfs it.
  Top right: Same as top-left, but now at 12.7\,d, which corresponds to the epoch of bolometric maximum
  (the spatial scale has changed).
   The dense ejecta wraps around the inner part of the disk. Near the shock, the ejecta is much hotter (bottom right quadrant),
   causing a local enhancement in radiative energy and flux (left quadrants).
   Bottom left: Same as top-left, but now at 115.7\,d, which corresponds to the time when the unshocked ejecta
   (i.e., ejecta material moving in a direction not intersecting the disk) becomes optically thin. The luminosity is from
   now on dominated by the ejecta/disk interaction.
   Bottom right: Same as top-left, but now at 231.5\,d.
   \label{diskmulti}
   }
   \end{figure*}

\begin{figure*}
    \epsfig{file=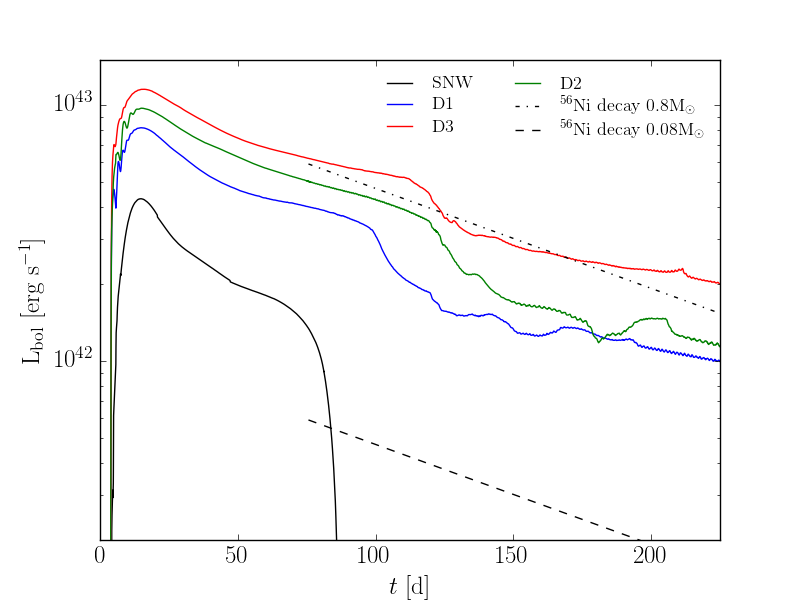,width=8.8cm}
    \epsfig{file=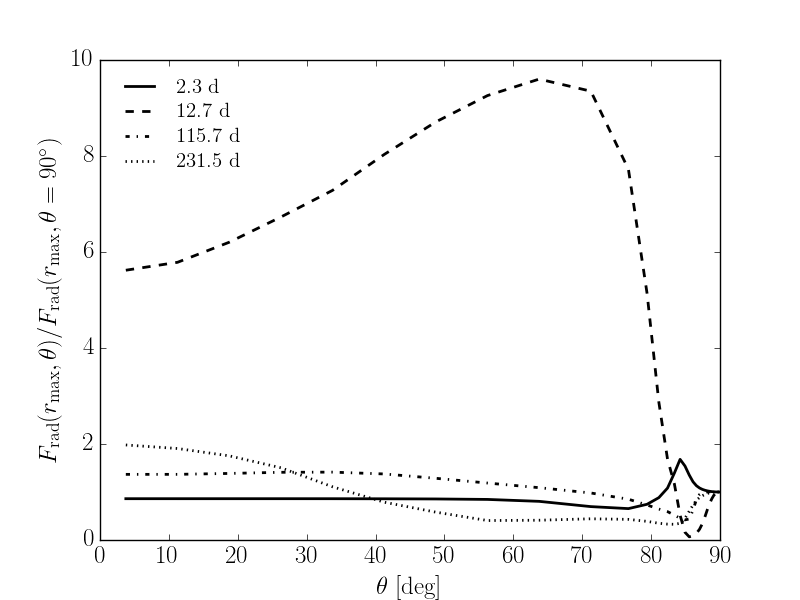,width=8.8cm}
    \caption{
    Left: Bolometric light curves for the ejecta-disk interaction models and
    the disk-less model counterpart (SNW). The radioactive decay power for an initial \iso{56}Ni mass of 0.08\,\msun\ (dashed line)
    and 0.8\,\msun\ (dash-dotted line) is overplotted.
   Right: Snapshots for model D1 of the emergent radiative flux $F_{\rm rad}(r_{\rm max}, \theta)$ versus polar angle,
   normalised to its equatorial value, and shown at 2.3, 12.7, 115.7, 231.5\,d after the onset of interaction.
    \label{lc_disk}
    }
\end{figure*}

\begin{figure*}
    \epsfig{file=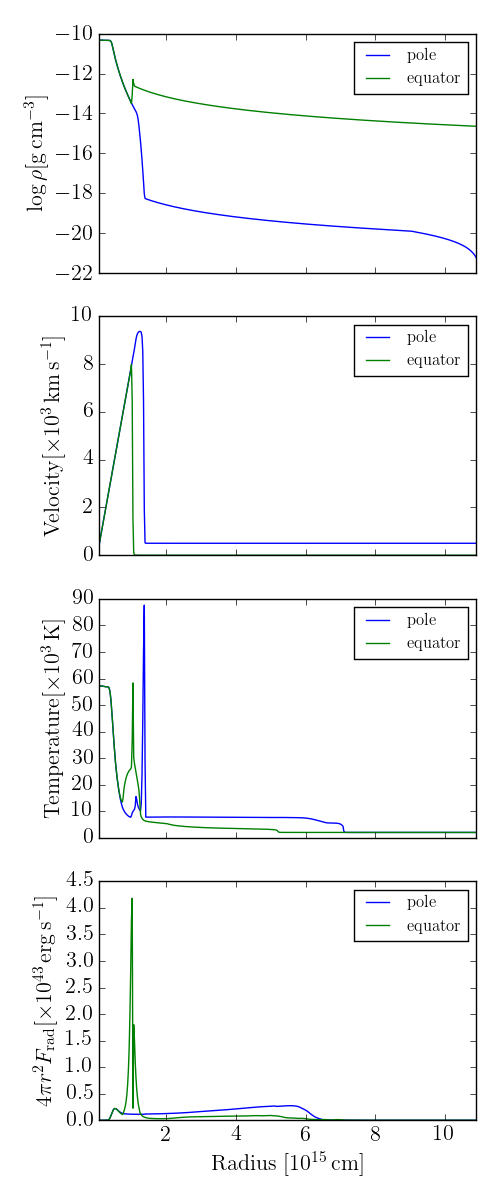,width=5.86cm}
    \epsfig{file=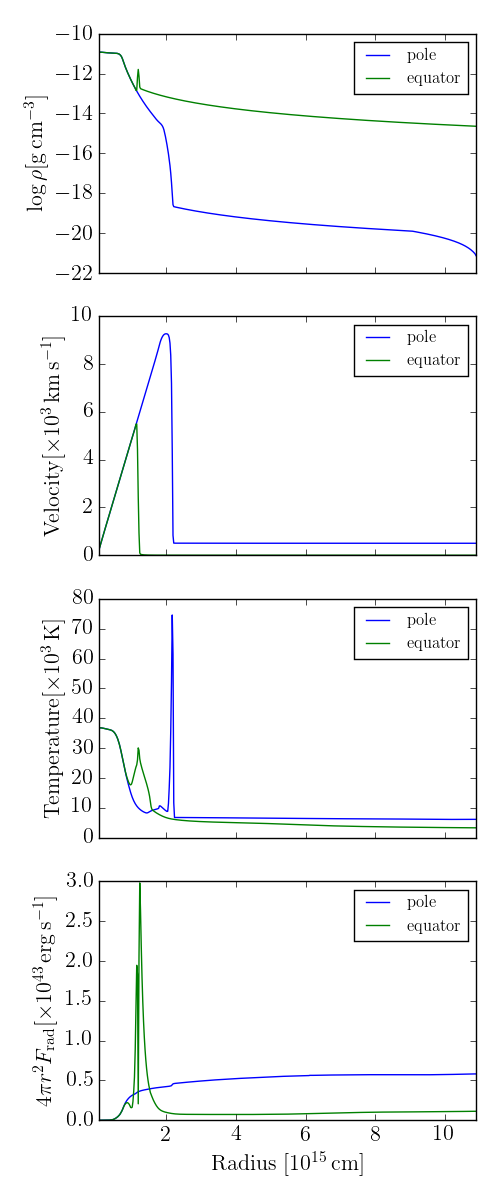,width=5.86cm}
    \epsfig{file=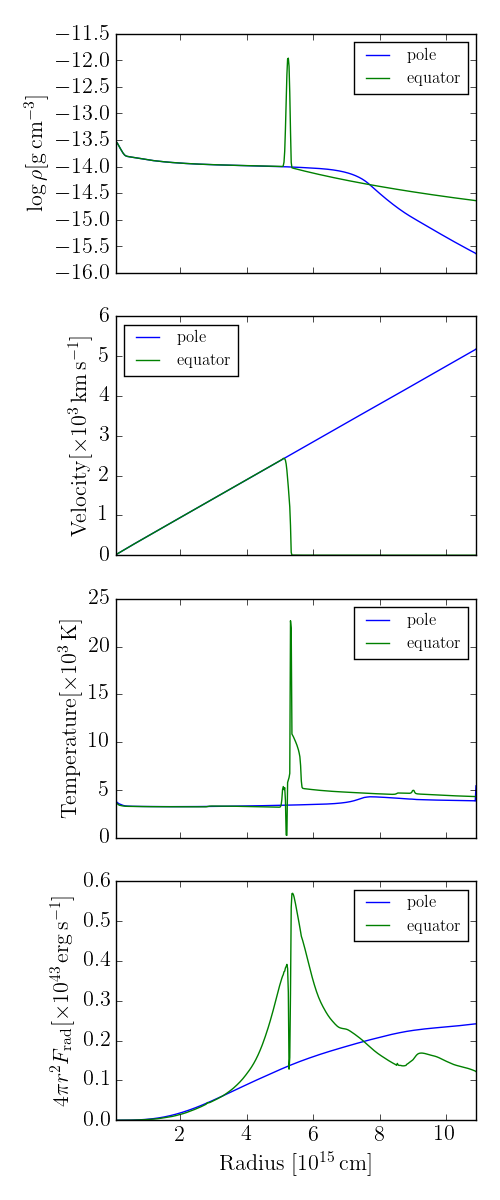,width=5.86cm}
    \caption{Radial variation of the density (top row), velocity (second row from top),
    temperature (3rd row from top), and total radiative flux (bottom row)
    along the polar and equatorial directions for model D5a
    at 2.3 (left), 12.7 (middle) and 231.5\,d (right) after the onset of interaction.
    The total radiative flux is scaled by $4 \pi r^2$ for easier comparison to a luminosity and to cancel the effect of spherical dilution.
    [See text for discussion.]
    \label{fig_D5a_cuts}
    }
\end{figure*}

We first discuss in detail the model D1. Figure~\ref{diskmulti} illustrates some properties of the ejecta/disk simulation
at 2.3, 12.7, 115.7, and 231.5\,d after the onset of interaction.
Figure~\ref{lc_disk} shows the evolution of the bolometric luminosity for model D1 and variants,
as well as the disk-less model counterpart (dashed line).
The bolometric luminosity  is used here to estimate the conversion efficiency of the ejecta/disk interaction
and is thus not the luminosity that a distant observer could infer.
Figure~\ref{fig_D5a_cuts} presents radial cuts at selected epochs for the velocity, density, temperature, and
radiative flux.

There are three regions to consider in this interaction:
the spherically-symmetric SN ejecta at radii $r<r_{\rm t}$; the disk
which lies along polar angles within 5\,deg of the equator; the material outside the disk and beyond the
outer edge of the ejecta. This material is low-density and does not influence sizeably the dynamics of the ejecta nor
the emergent radiation. Hence, the regions of interest are the $\sim$10\,d-old SN II ejecta, which radiates
its stored energy (in isolation, it contributes the bolometric light curve tagged ``SNW'' in Fig.~\ref{lc_disk}),
and the ejecta fraction that interacts with the disk. The radial velocity of the disk material is zero so all
ejecta mass shells move faster than the disk material. The fraction of the ejecta that will interact with the disk
is thus given by the fractional solid-angle subtended by the disk, which is equal to $\sin \theta_{\rm D}$.
In model D1, this corresponds to $\lesssim$\,9\% --- that value also gives the maximum conversion efficiency
of the ejecta/disk interaction in model D1.

In the top-left panel of Fig.~\ref{diskmulti}, we see the properties of the system at 2.3\,d. The radiation from the SN
ejecta fills up the grid; the photosphere is located around 10$^{15}$\,cm. The interaction along the equator
is strong and causes a large local enhancement of the flux. Some of this radiation escapes, and the fraction that
is trapped raises the radiative energy and the gas temperature in the vicinity (top-left and bottom-right quadrant).

At 12.7\,d (top-right panel in Fig.~\ref{diskmulti}), the CDS that forms at the ejecta/disk interaction site moves
very slowly, so that the faster SN ejecta material engulfs this interaction region.
The energy released by the shock cannot escape because the surrounding material is optically thick so
the ejecta regions in the vicinity of the shock heat up (compare with the ejecta regions at the same $r$ but along
the pole). The material within the disk and ahead of the CDS heats up too and the photosphere along the equator is now at a radius
nearly twice as large as its value along the pole. The emergent flux varies strongly with latitude
(right panel of Fig.~\ref{lc_disk}).
It is minimum along the equator where the radial optical depth is maximum
(($F_{\rm p}/F_{\rm e})_{r_{\rm max}}=$\,5.1). The emergent flux is maximum along the edge of the disk
(($F_{\rm p}/F_{\rm e})_{r_{\rm max}}=$\,9.0), showing contributions from the ejecta
and from the ejecta/disk interaction. Progressing towards
the pole, the emergent flux decreases because the contribution from the interaction diminishes.
In a cumulative sense, the interaction sizeably enhances the luminosity from the system compared to the disk-less model counterpart,
by about a factor of two at maximum (Fig.~\ref{lc_disk}).

At 115.7\,d (bottom-left panel in Fig.~\ref{diskmulti}), the SN ejecta is nearly entirely optically thin (recall that we ignore
any contribution from radioactive decay, which would merely lengthen the optically-thick phase). Within the disk,
only the locations close to the CDS are optically thick. The CDS velocity has now dropped from 10000\,\kms\ initially
to 2700\,\kms. Importantly, the disk material is at rest so the interaction will continue as long as there is disk material --- the
asymptotic velocity of the CDS could be zero.
The only source of luminosity at that time is the ejecta/disk interaction, which gives a slow-decreasing tail to the light
curve (Fig.~\ref{lc_disk}).

At 231.5\,d (bottom-right panel in Fig.~\ref{diskmulti}), the interaction is still going.
The latitudinal variation of the flux is complex, with the two large contributions arising
from the equator and the pole (with ($F_{\rm p}/F_{\rm e})_{r_{\rm max}}=$\,2) and a flux deficit in between.
The excess along the polar direction probably arises from the axial-symmetry and the radiation of the disk
in the equatorial plane, unobscured by the SN ejecta material which is cold and recombined.

To illustrate more clearly the above discussion, we show in Fig.~\ref{fig_D5a_cuts} the radial variation
of the velocity, density, temperature, and radiative flux along the pole and the equator at 2.3 (left), 12.7 (middle)
and 231.5\,d. This figure shows the velocity profile, which is absent in the previous colormaps.

Compared to the disk-less model counterpart, model D1 exhibits a greater bolometric luminosity
at all times. The interaction with the disk yields a factor of two increase in the total luminosity
during the high brightness phase (which corresponds to the phase during which the SN ejecta is
optically thick). The extra heat from the interaction also enhances the ejecta ionisation, and thus its opacity,
lengthening the high-brightness phase by about 20\%.
As the ejecta turns nebular, the ejecta/disk interaction continues to power the light curve, at a rate that
is comparable to that of $^{56}$Co decay but for an original $^{56}$Ni mass of 0.44\,\msun.
Such a $^{56}$Ni mass is well above the value inferred for standard core-collapse SNe so this teaches
us two things.
First, if a disk-ejecta interaction is invoked to produce a super-luminous SN IIn, the disk mass/thickness
required  to boost the high-brightness phase of the SN will also yield a huge nebular luminosity.
Second, a disk-ejecta interaction could power the nebular flux of a SN II but to mimic the effect
of 0.08\,\msun\, of  $^{56}$Ni, it would also have a negligible effect at earlier times.

\subsection{Results for different disk mass and thickness}

  We now discuss how the results for model D1 change when we increase the disk mass from 1.5 to 5.0\,\msun\ (same
  disk opening angle; model D2) and when we double the disk opening angle (same disk mass; model D3).
  The evolution of these two additional interaction models is comparable to that of model D1 so we focus
  on the differences in the bolometric luminosity (left panel of Fig. \ref{lc_disk}).

   Both models D2 and D3 show a larger bolometric luminosity at all times compared to model D1.
   By increasing the disk mass for the same disk opening angle in model D2 (which therefore enhances
   the disk density), we enhance the ejecta deceleration and the conversion efficiency of the interaction.
   By increasing the disk opening angle in model D3, we enhance the fraction of the ejecta that takes
   part in the interaction. Here, by doubling the opening angle, we come close to doubling the luminosity offset between
   model D1 and the disk-less model counterpart (the gain is not quite a factor of 2 because the disk mass was kept the
   same, hence the disk density was halved).

    To turn a standard SN II into a super-luminous SN IIn by means of a disk-ejecta interaction, as proposed
    by \citet{Metzger2010}, is possible but it requires a disk
    with a very large opening angle (enhancing the mass/density of a thin disk will not work because a too
    small fraction of the ejecta interacts with the disk). The other issue is that during the
    high-brightness phase, the ejecta-disk interaction is obscured, and one expects to see a dominance of
    broad lines from the (unshocked) ejecta. The maximum-light spectra for such an event would therefore not
    show narrow lines and would not be typical of SN IIn.

    Interaction with a thin and moderate mass disk is, however, a possible power source for the nebular flux
    in a core-collapse SN, as proposed by \citet{Smith2015_PTF}.

\section{Conclusions}
\label{sec_discussion}

In this work we have investigated the dynamical and radiative properties of ejecta/CSM interactions
by means of axially-symmetric (2-D) multi-group radiation-hydrodynamics simulations with the code \heracles.
We have covered a variety of interaction configurations, including symmetric ejecta and asymmetric wind CSM, asymmetric
ejecta and symmetric wind CSM, and finally symmetric ejecta and disk CSM.
In our approach, we break the symmetry in the initial model by introducing a latitudinal scaling of the density distribution
in the ejecta or in the CSM.
Our work extends the preliminary 2-D calculations of SNe IIn by \citet{van_marle_etal_10}. Their approach
does not solve for the radiation field,  but instead treats the  radiation through a parametrised optically-thin cooling function,
which ignores the trapped radiation energy and its
diffusion through the unshocked CSM and thus does not capture the strong optical-depth effects inherent to super-luminous
SNe IIn.

Interactions involving an asymmetric wind CSM produce a depth and latitude-dependent radiative flux.
Along directions where the CSM density is larger, the ejecta deceleration is greater and the shock
luminosity is enhanced. But because the radial optical depth is also greater along those angles,
the emergent flux is reduced. Instead, the flux emerges from regions of lower density, which correspond
to directions in which the deceleration is weaker.
This optical-depth effect (and associated redistribution in angle) persists as long as the CSM optical
depth at the CDS is large, a situation that can last for months in some super-luminous SNe.
The asymmetric distribution of mass in the CSM also affects the motion of the CDS. An oblate CSM
density distribution yields a prolate CDS.
During the evolution of the system, the CDS is initially located deep below the photosphere in the CSM.
Eventually, the photosphere is contained in the CDS. For a prolate CSM density distribution, the
photosphere is thus prolate early on but becomes oblate at late times (this could for example
produce a 90-deg flip in the polarisation angle).
Our simulations provide a useful framework for the interpretation of the polarisation observed in super-luminous
SNe IIn  \citep{Patat_2011}. During the high brightness phase, optical-depth effects are strong and yield
a distribution of the flux that is anti-correlated with the distribution of scatterers --- this result
was already discussed in the radiation-transfer simulations of \citet{Dessart_2011} and \citet{Dessart_2015}.
The complexity of the system is however a challenge for a robust interpretation of the polarisation
at late times.

Interactions involving an asymmetric ejecta (but a symmetric wind CSM) tend to produce a smaller
latitudinal variation in the emergent flux because now the optically-thick CSM damps
the variations of the radiation injected at the shock (which may itself be very aspherical).
Even for a strongly asymmetric explosion, the latitudinal variation of the emergent flux is negligible
at bolometric maximum if the shock is embedded within a (spherically-symmetric) CSM whose
electron-scattering optical depth is $\gtrsim$\,10.
However, at later times, the optical depth at the shock/CDS is smaller and a strong angle-dependence
in the emergent flux can be seen.
Given this, the observation of a sizeable polarisation at maximum light in SN\,2010jl \citep{Patat_2011}
supports an asymmetry of the CSM, as proposed by D15.

In section \ref{sec_disk} we examined the interaction between a SN ejecta and a relic disk located at 10$^{15}$\,cm.
We considered disks of various opening angles (5 and 10\,deg) and masses (1.5 and 5.0\,\msun).
The ejecta/disk interaction boosts at all times the luminosity of the model counterpart without a disk.
During the high brightness phase of $\sim$\,100\,d, the boost stems from shock deposited energy
within the optically-thick ejecta, also introducing a strong angle dependence to the flux.
During the nebular phase, the luminosity is dominated by the ejecta/disk interaction. Interaction with
a disk of small mass/thickness can easily exceed the decay power observed in standard (non-interacting)
type II SNe --- this scenario may apply in some SNe \citep{Smith2015_PTF}.
However, producing a super-luminous SN II through an ejecta/disk interaction \citep{Metzger2010}
requires a massive disk with a very large opening angle in order to tap a large fraction of the ejecta
kinetic energy. Given that most of the ejecta would not be interacting, this model would unlikely
produce a narrow-line spectrum during the high-brightness phase and thus does not seem adequate
for a SN IIn.

\section*{Acknowledgments}

AV and LD acknowledge financial support from ``Agence Nationale de la Recherche"
grant ANR-2011-Blanc-SIMI-5-6-007-01.
This work used the computing resources of the M\'esocentre SIGAMM,
hosted by the Observatoire de la C\^ote d'Azur, Nice, France.

\label{lastpage}

\end{document}